\documentclass[fleqn,usenatbib]{mnras}

\usepackage[T1]{fontenc}
\usepackage{ae,aecompl}

\usepackage{newtxtext,newtxmath}

\usepackage{graphicx}	
\usepackage{amsmath}	
\usepackage{tablefootnote}	

\title[Galactic star formation history]{
	Anatomy of galactic star formation history: Roles of 
	different modes of gas accretion, feedback, and recycling
}

\author[M. Noguchi]{
Masafumi Noguchi $^{1}$\thanks{E-mail: noguchi@astr.tohoku.ac.jp (MN)}
\\
$^{1}$Astronomical Institute, Tohoku University, 6-3, Aramaki, Aoba-ku, Sendai, Miyagi, 980-8578, Japan
}

\date{Accepted XXX. Received YYY; in original form ZZZ}

\pubyear{2020}

\begin{document}
\label{firstpage}
\pagerange{\pageref{firstpage}--\pageref{lastpage}}
\maketitle

\begin{abstract}

	We investigate how the diverse star formation histories observed across 
	galaxy masses emerged  using  
	 models that evolve under gas accretion from host halos.
	They also include 
	ejection of interstellar matter by supernova feedback, recycling of ejected matter
	and preventive feedback that partially
	hinders gas accretion.
	We consider three schemes of gas accretion: the fiducial scheme which includes
	the accretion of cold gas in low-mass halos and high-redshift massive halos as hinted by
	cosmological simulations; 
	the flat scheme in which high-mass cold accretion is removed; and finally   
	the shock-heating scheme which assumes radiative cooling of the shock-heated halo gas.
	The fiducial scheme reproduces dramatic diminishment in star formation rate (SFR) 
	after its peak as observed for the present halo mass 
	$M_{\rm vir}>10^{12.5}{\rm M}_\odot$ while other two schemes show reduced or negligible 
	quenching.
	 This scheme  reproduces the high-mass slope in the SFR vs. stellar mass relation 
	 decreasing toward recent epochs whereas other two schemes show opposite trend which 
	 contradicts observation.
	 Success in the fiducial scheme originates in the existence of high-mass cold-mode accretion 
	 which retards transition to
	 the slow hot-mode accretion thereby inducing a larger drop in SFR.
	 Aided by gas recycling, which creates monotonically increasing SFR in low-mass halos,
	 this scheme can
	 reproduce the downsizing galaxy formation.
	 Several issues remain, suggesting non-negligible roles of missing physics. 
	 Feedback from active galactic nuclei could mitigate 
	 upturn of  SFR in  low-redshift massive halos
	 whereas galaxy mergers could remedy early inefficient star formation.

\end{abstract}


\section{introduction}

Recent large scale surveys both from the ground and the space 
(e.g., SDSS, GALEX, PRIMUS, CANDELS, ZFOURGES)
have revealed the properties of galactic star formation
across a  wide mass range form the local Universe to high redshift
utilizing multiwavelength observations
\citep[e.g.][]{ma05,co11,gr11,ko11,sk14,st16}.
The most important information derived from these surveys includes 
the star formation history (SFH) for various masses and the correlation between 
star formation activity measured by the star formation rate (SFR) and other properties 
of host galaxies.

Many theoretical attempts have been made to reproduce and interpret the data 
obtained by the large surveys thereby understanding the galaxy formation process over the cosmic time.
They aim to explain not only the observational data directly related to the star formation activity but also other global properties
of galaxies such as luminosity and mass functions, colour distributions, sizes, metallicity, gas contents,
scaling relations (e.g., Tully-Fisher relation).
They are
broadly classified into two categories, namely, cosmological
simulations and semi-analytic models.

Cosmological simulations (e.g., Illustris, EAGLE, FIRE, NIHAO, VELA) 
follow the evolution of dark matter halos and galaxies in the Universe
 starting from the initial condition prescribed by the assumed cosmology.
 Growth of dark matter halos and several baryonic processes (e.g., gas heating and cooling on global scales)
 are directly treated. Due to the limited spatial and time resolution attainable, other local processes
 such as star formation, stellar evolution, feedback, black hole growth are usually 
 treated as 'subgrid' physics.
Within this limitation, cosmological simulations
are expected to provide the most realistic and direct treatment for galaxy
formation. 
However, the results depend strongly on the prescription for the subgrid physics
and much effort has been paid to understand the nature of such a dependence and improve the modelling
\citep[e.g.][]{ce09,ho12,ge14,ho14,vo14a,sy15,wa15}.
Although these simulations provide a powerful tool in
making realistic galaxies, their expensive nature sometimes makes it difficult to
elucidate the effect of included processes by running a large suit of models.

Semi-analytic models (e.g., GALFORM, L-GALAXIES, GalICS 2.1) play a complementary role 
to the cosmological simulations.
They build on the hierarchical merger history of the dark matter halos taken from
the analytical or numerical studies and include various physical processes
that govern the evolution of baryonic components
   \citep[e.g.][]{wh91,ka93,la93,co94,so99,cr06,mo07,de07,de10,gu11,la14,lu14,la16,cr16,mi18,ca20}.
These processes are usually included in parametrized forms and effects of these parameters 
are investigated. This approach has an advantage that it can explore wide ranges in parameters 
which is usually difficult for expensive cosmological simulations.
The drawback is that the assumptions about baryonic processes affect result more 
strongly than in cosmological simulations and cannot handle spatial configurations such as
galaxy morphology directly.

Currently, these two approaches, sometimes in cooperation, are attaining remarkable success in reproducing a variety
of observed characteristics for wide ranges in galaxy masses and cosmological epochs.
However, these two methods  have a drawback in common that
the complicated interplay of employed physical processes poses obstacles to clear interpretation 
of the obtained results in some cases although this difficulty is partially overcome by running 
models including different sets of physical processes.

Several studies (e.g., UNIVERSE MACHINE, EMERGE) utilize the abundance matching method and other similar 
techniques to connect the merger history of 
dark matter halos predicted by numerical simulations 
and the observational data for galactic star formation \citep[e.g.][]{be13,be19,mo13,mo18,mo21}.
In short, those studies parametrize the baryonic processes such as the gas accretion rate, the star formation rate, 
and the gas ejection rate
 as a function of halo mass and redshift and the parameters are tuned so as to
reproduce the observational data. Due to their made-to-measure nature, these approaches are regarded as giving observational
constraints rather than theoretical prediction. Nevertheless, they enable linking galaxies at 
different epochs into continuous evolutionary paths empirically and serve as a valuable testbed for theoretical 
models. The results from these studies are intensively used in our study for comparison.

We take here a new approach using a simple evolution model of galaxy evolution
that includes a minimum set of potentially important processes. By intercomparing models incorporating different 
gas accretion schemes and different sets of feedback and gas recycling, we aim 
to isolate the effect of each ingredient.
Our modelling is thus in a similar spirit to the 'equilibrium' model of galaxy evolution
\citep[e.g][]{da12,mi15}
but does not assume the instantaneous equilibrium of the galactic system.
It traces the time evolution of galaxies explicitly without the constraint of equilibrium
and is expected to give more realistic treatment. For example, the evolution of 
gaseous contents in galaxies, set automatically constant with time in the equilibrium model,
can be directly traced and compared with the observational data.
Mass loading factor for stellar (supernova) feedback is also obtained automatically.
Early phase of galaxy evolution driven by efficient gas fuelling 
likely deviates from the equilibrium condition \citep{da12}.
This phase is an important target of this study so that we think the freedom from equilibrium assumption
is a rewarding feature of the present model.
Because our model nevertheless involves a number of simplifications, we hope to reproduce 
qualitative features of galactic star formation gained by the observation in the first place.
We will see, however, that the present model gives quantitatively reasonable description of the 
observation in some cases.
Many previous studies put forward propositions for the respective roles of the 
physical processes involved in the evolution of stellar contents of galaxies.
By constructing the model that satisfies a wide range of observational constraints simultaneously, we try to
check these propositions and get a new insight into key physics that governs galactic star formation history.

Our paper is organized as follows. In section 2, we introduce the models we use for tracing galaxy evolution.
Section 3 describes and discusses the gas accretion schemes explored in this study. Star formation histories 
in these schemes are examined in detail and compared with observations in section 4. Section 5 is devoted 
to the investigation of roles played by the physical processes included in the model. Section 6 presents
discussion on several issues for which further study is desired and section 7 summarizes our conclusions.

\section{Models}

\begin{figure*}
	\includegraphics[width=\linewidth]{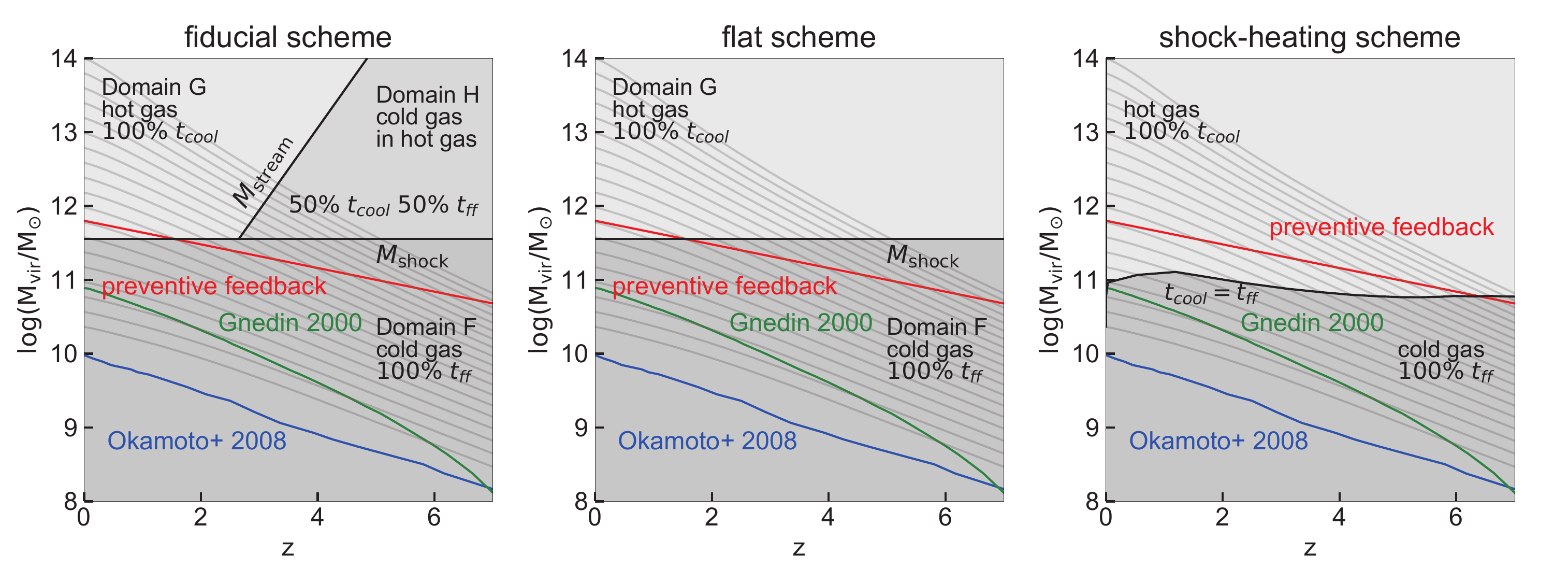}
    \caption{
	    Three schemes for the gas accretion on to discs
	    illustrated in the ($z-M_{\rm vir}$) plane.
	For the fiducial scheme (left), three domains exist bordered by 
	the shock mass, $M_{\rm shock}$, and the stream mass, $M_{\rm stream}$, 
	as explaind in the text. For the flat scheme (centre), the domain H is 
	removed. The domains in the shock scheme (right) are separated by the border
	for which the cooling time is equal to the freefall time.
	In each domain, the halo gas accretes to the disc with the free-fall time  ($t_{\rm ff}$)
	or the radiative cooling time ($t_{\rm cool}$) as indicated.
	The green line indicates the filtering mass proposed by \citet{gn00}, 
	whereas the blue line indicates the virial mass below which the  fraction 
	of the baryon retained by the halo (relative to the cosmic baryon fraction)
	decreases to less than 90 percent due to UV background radiation from \citet{ok08}. 
	Red line indicates the characteristic mass for preventive feedback as explained in the text.
	Grey lines show the evolution track of each halo.
	}
\end{figure*}

We model a disc galaxy as a composite system comprising the halo gas, the interstellar gas, and the 
stellar disc embedded in a dark matter halo. The baryonic components (the halo gas, the interstellar 
gas, and the stellar disc) are specified by their masses, the time evolution of which is calculated 
over the cosmic time. This model is thus 'one-zone' treatment in the sense that the spatial structure of
each baryonic component (and the dark matter halo) is not considered except when we calculate 
the rates of several physical processes as explained later. 
The physical processes included in our model are the growth of the 
dark matter halo, the accretion of the halo gas on to the disc (i.e., 
creation of the interstellar gas), the star formation from the interstellar gas, the ejection of the 
interstellar gas by supernova feedback, the reincorporation (recycling) of the ejected gas 
to the interstellar gas, and the prevention of the gas accretion on to the disc.
We describe the physical aspect of those processes here 
and defer numerical details of their implementation to Appendix.
We assume the $\Lambda$CDM cosmology with parameters $\Omega_M$=0.27, $\Omega_\Lambda$=0.73,
h=0.7, $n_s$=0.95, and $\sigma_8$=0.82.

\subsection{Dark matter halo growth}

We take the parametrized form for the mass growth of the dark matter halo given 
in equations (H1)-(H6) in \citet{be13}, which is tested against Bolshoi, 
Consuelo, and Multidark cosmological simulations.

\subsection{Gas accretion}

Gas accretion on to the galactic discs and subsequent star formation from 
the interstellar gas is
the most important driver of the galaxy formation and evolution.
In recent development of the hydrodynamical cosmological 
simulations, we have 
seen a significant revision of the traditional view about the physics of cosmic gas accretion.

The widely-used scenario for gas accretion on to forming galaxies, namely, the shock heating 
paradigm, assumes that the gas entering a dark matter halo is heated by 
shock waves to roughly virial temperature of the halo and subsequent radiative cooling
induces the accretion of this gas to the inner parts where the galaxy grows
\citep[e.g.][]{re77}. In this picture, the gas accretes with the free-fall time if 
it cools rapidly while the accretion proceeds with the radiative cooling time 
when the cooling is not efficient. The cooling argument underlying this scenario succeeded in explaining 
the segregation 
of galaxies and galaxy clusters on the density-temperature plane \citep{bl84} and has constituted 
a backbone for many theoretical studies.

However,  realistic simulations for
hydrodynamical evolution of the primordial gas in the $\Lambda$CDM Universe 
 \citep[e.g.][]{fa01,ke05,de06,oc08,va12,ne13}
  found that the assumption of the stable shock underlying this picture is not always correct.
  The shock wave is unstable and no shock heating occurs in low mass halos.
  The unheated cold gas directly accretes to the inner parts in free-fall fashion in this case
  \citep[e.g.][]{fa01,bi03,ke05,de06}.
  In addition to this cold flow in low mass halos, high mass halos at high redshifts 
  develop filaments of cold gas inflowing through the surrounding hot gas created by shocks
  \citep[e.g.][]{de06,oc08,va12,ne13}.
 The new picture, sometimes called the cold accretion scenario, 
   has a large impact on our understanding of galaxy evolution,
   For example, it helps interpret the high abundance of actively star-forming galaxies at
   high redshifts
     \citep[e.g.][]{ca06,de09}. 
     Lyman limit systems may be observational manifestations of the predicted cold streams \citep{fu11}.
     The  Ly$\alpha$ emitting filaments at high redshift observed by \citet{ma16} may provide a more direct example.
    Extending this picture to larger spatial scales, 
    \citet{ov16} proposes that the transition from the cold accretion regime to the regime 
    dominated by the hot intracluster gas in most massive halos marks the division between protoclusters and clusters of galaxies.

\citet{no18} demonstrated that the cold accretion scenario, combined with 
a chemical evolution model, reproduces the bimodality in the
stellar abundance distributions of the Milky Way disc stars
(for new observations, see \citet{qu20}).
 The transition from the early cold-mode to the late hot-mode gas accretion 
 sets the stage for
 two distinctive star formation episodes (see also \citet{bi07}). The short period active star formation in early times
 produces metal-poor (i.e., deficient in Fe) stars with relatively enriched $\alpha$-elements by the dominant 
 action of type II supernovae.
The late-time  star formation proceeds slowly for a longer period allowing much iron enrichment by
type Ia supernovae, leading to a metal-rich $\alpha$-element-poor (low $[\alpha/{\rm Fe}]$)
population.
Several cosmological simulations also show signs of such two-stage star formation in 
Milky Way mass galaxies
\citep[e.g.][]{ho14,gr18}, although details are often masked by other baryonic processes included.
For galaxy morphology, the cold flow scenario was found to reproduce the relative dominance of
thin discs, thick discs, and bulges in the present-day disc galaxies as a function of galaxy mass
by associating the hot mode with thin  disc formation and the cold mode with bulge and
thick disc formation, respectively \citep{no20}. The observed galaxy-mass dependence 
of the mean age and the age spread of the bulge \citep{br18,br20} is also naturally explained by this model.
We adopt this cold flow picture in our fiducial scheme described later and examine its essential role by
comparing it with the accretion scheme based on the shock-heating picture.

\subsection{Star formation}

The accreted halo gas turns into interstellar gas in the galactic disc and serves
as raw material for star formation.
To calculate the star formation rate (SFR), we adopt the formulation by \citet{bl06} in which 
the SFR is essentially determined by the molecular gas content in the interstellar medium (ISM),
with the molecular fraction in turn controlled by the disc midplane pressure.
This treatment enables us to trace the evolution of the atomic and molecular contents
separately. Those results are compared with the observation.
Mass-loss associated with the evolution of massive stars is evaluated by assuming 
Chabrier IMF with lower and upper initial mass limits set to 
$1 {\rm M}_\odot$ and $100 {\rm M}_\odot$, respectively. 
The resulting mass fraction, $R=0.44$ is returned to the ISM instantaneously (see Appendix A3).

\subsection{Feedback}

It is well established that the evolution of low-mass galaxies is strongly affected by
the ejection of the ISM by supernovae on both observational and theoretical grounds 
\citep[e.g.][]{de86,ma99,br07,mc19}. It not only helps explain observations such as 
the mass-metallicity relation but also solves the over-production problem in the low galaxy mass domain inherent in 
the $\Lambda$CDM paradigm (see \citet{bu17} for a review and \citet{we12} for detailed  discussion).
We take the energy-driven feedback as the standard choice although the model with the
momentum-driven feedback is also discussed.

Recent studies \citep[e.g.]{to14,so15} suggest that supernova feedback alone cannot produce the masses of low-mass
galaxies consistent with their metallicity.
In order to realize the observed metallicity, those galaxies should be more massive than observed.
Another type of feedback, called preventive feedback, was found to resolve this tension. 
Preventive feedback refers to the suppression of gas accretion on to halos or galactic discs by 
some mechanism(s). It can thus reduce the stellar mass of these galaxies without 
affecting metal enrichment driven by ejective feedback.  
\citet{lu17} point out the importance of preventive feedback in reproducing the galaxy stellar mass function and 
the mass-metallicity relation of the Milky Way satellites simultaneously.
Several cosmological simulations  \citep{an17,ch16,op10} show that this type of preventive feedback 
indeed occurs in galaxy formation although its detailed nature is not completely clarified.
 Candidate mechanisms include radio-mode feedback by active galactic nuclei (AGNs) \citep{ke09}, heating from 
 outflowing winds \citep{op10}, and heating by ionizing radiation field \citep{ef92,gn00,ho06,ok08,ch16}.
Preventive feedback 
can take two different forms. One is the reduction of the gas entering 
dark matter halos and another is the suppression of the gas supply from parent halos to 
galactic discs. Although this should be discriminated in the strict sense \citep[e.g.][]{ch16}, we include this 
process by reducing the accretion of the halo gas to the disc by a specified factor.
We confirmed that the reduction of the gas entering the halo by the same factor
gives indistinguishable results. 

\subsection{Gas recycling}

Outflow of the interstellar gas is ubiquitously observed in star-forming galaxies with a large mass range
and taken as a signature of ejective feedback.
Because the outflow velocity attained  by supernova feedback is typically less than several 
hundred kilometre per second
\citep[e.g][]{ch15},
the outflowing gas may not leave the host galaxy entirely but a part of it may eventually fall back to
the galaxy.
Many numerical simulations demonstrate that such a fall-back actually occurs in actively star-forming 
galaxies 
\citep[e.g.][]{ch16,gr19}.
This type of gas recycling has also been incorporated in the semi-analytic models (SAMs) \citep[e.g.][]{cr06,he13,wh15}
and its effect on galaxy evolution has been intensively studied.
These studies show that recycling affects not only global properties of galaxies but also their 
internal structures 
\citep[e.g.][]{br12,ma12,gr19}.
On the other hand, observational confirmation of gas recycling turned out to be inconclusive
(see \citet{fr18} for a review).
Although the infall of cold gas like galactic fountains is observed, several possibilities 
can be considered for its origin, including the cooling flow of the hot halo gas or the inflowing 
streams of unheated gas (cold-mode accretion). It is theoretically anticipated that 
the circulation of the ejected gas over much larger scales than the observed local
fountains also takes place \citep[e.g.][]{an17,mi20}. Recent studies \citep[e.g.][]{fo14,ha22} suggest that 
the recycling gas may be observed as quasar absorption lines (especially ${\rm C}_{\rm IV}$)
but more work is required to firmly establish their connection. Considering these situations, we include this 
process simply by returning the ejected gas to the disc component with a specified timescale.

\begin{figure*}
	\includegraphics[width=1.0\linewidth]{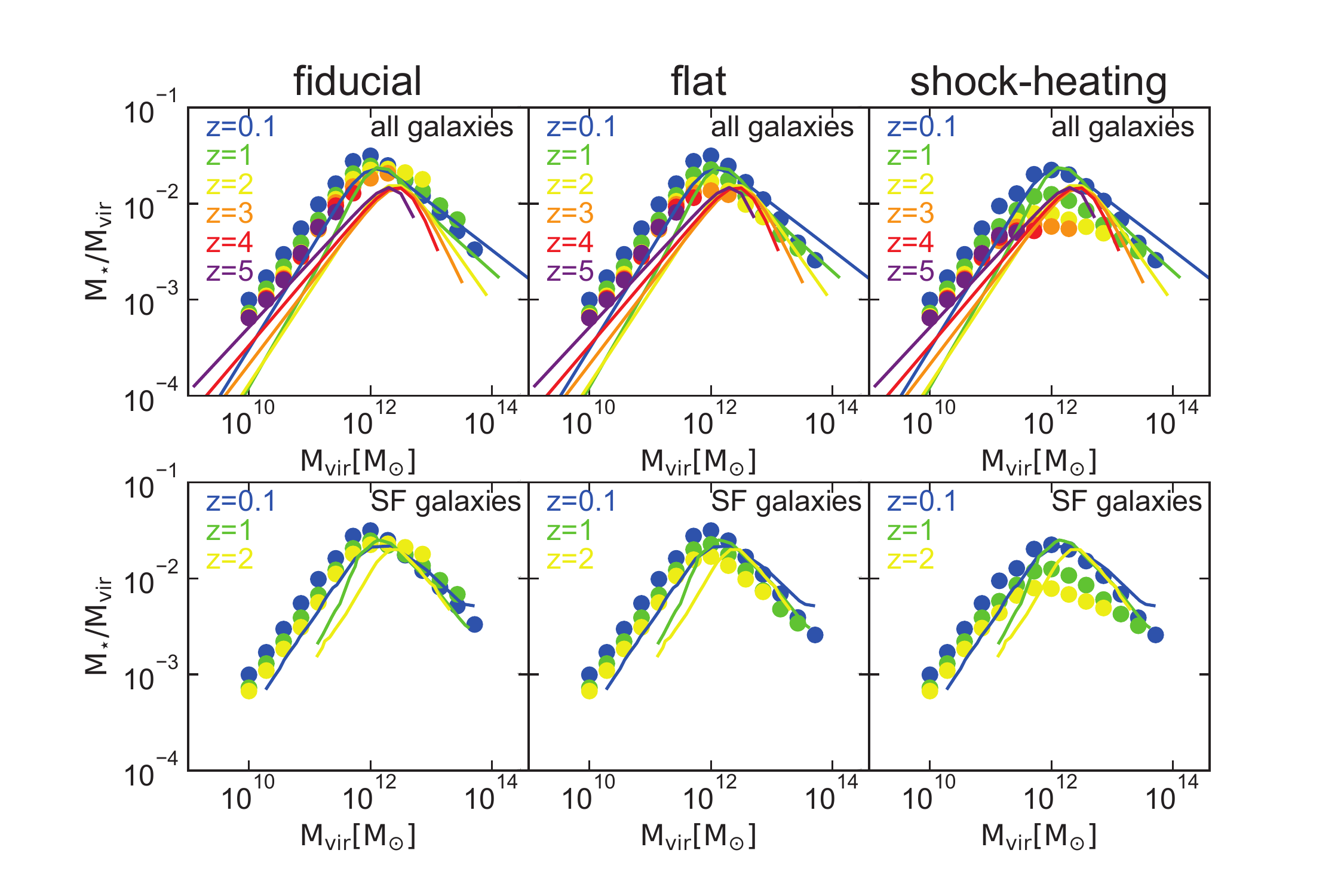}
    \caption{
	    The ratio of the stellar mass to the halo virial mass as a function of 
	    the virial mass and redshift. Filled circles indicate the result 
            for the schemes considered here. The upper panels compare the model with
	    the results for all galaxies (quiescent and star forming) by \citet{be19}, 
	    whereas the lower panels 
	    compare the model with the result by \citet{be19} only for star-forming 
	    galaxies.
         }
\end{figure*}

\begin{figure*}
	\includegraphics[width=1.0\linewidth]{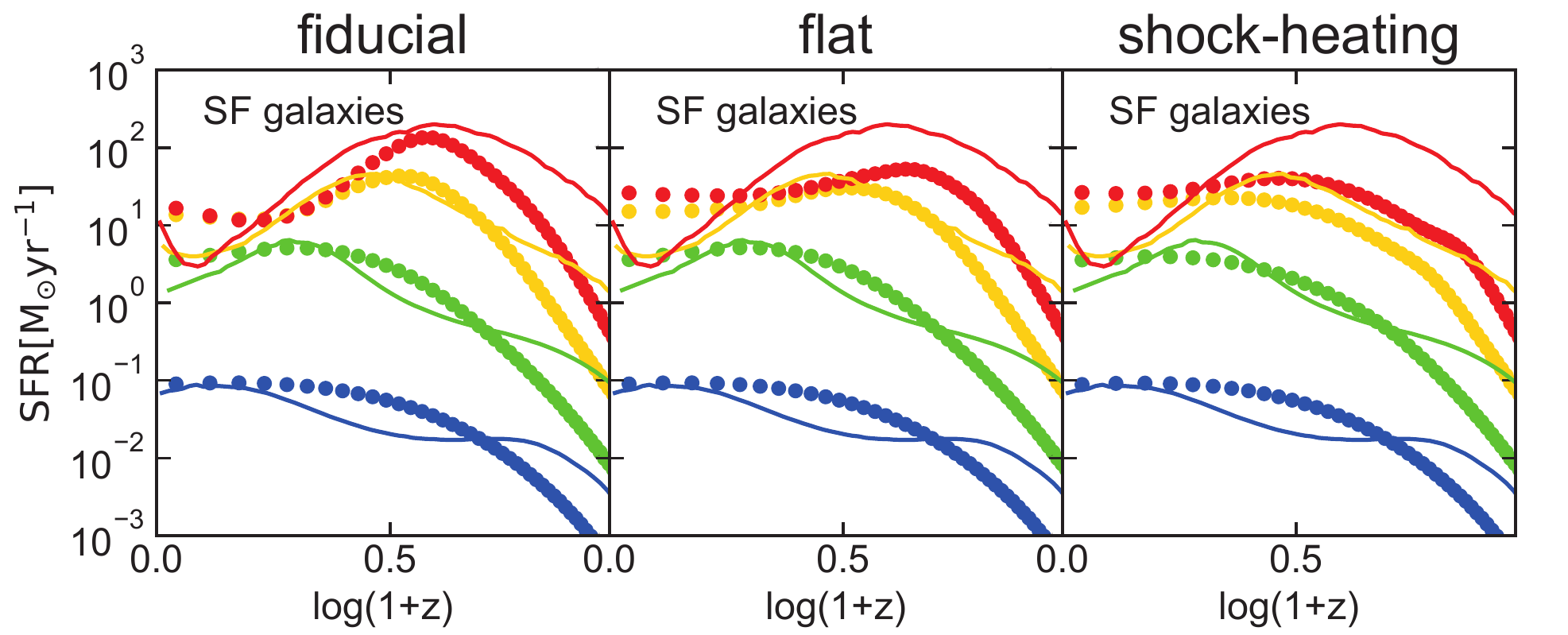}
    \caption{
	    Star formation history as a function of the present halo virial mass.
	    Filled circles with different colours denote SFR for different values for 
	    the present halo mass (blue: $10.5<{\rm log} M_{\rm vir}<11.5$,
	    green: $11.5<{\rm log} M_{\rm vir}<12.5$,
	    yellow: $12.5<{\rm log} M_{\rm vir}<13.5$,
	    red: $13.5<{\rm log} M_{\rm vir}<14.5$).
	    Solid lines show the results from \citet{be19} for star-forming 
	    galaxies. 
    }
\end{figure*}

\begin{figure*}
	\includegraphics[width=1.0\linewidth]{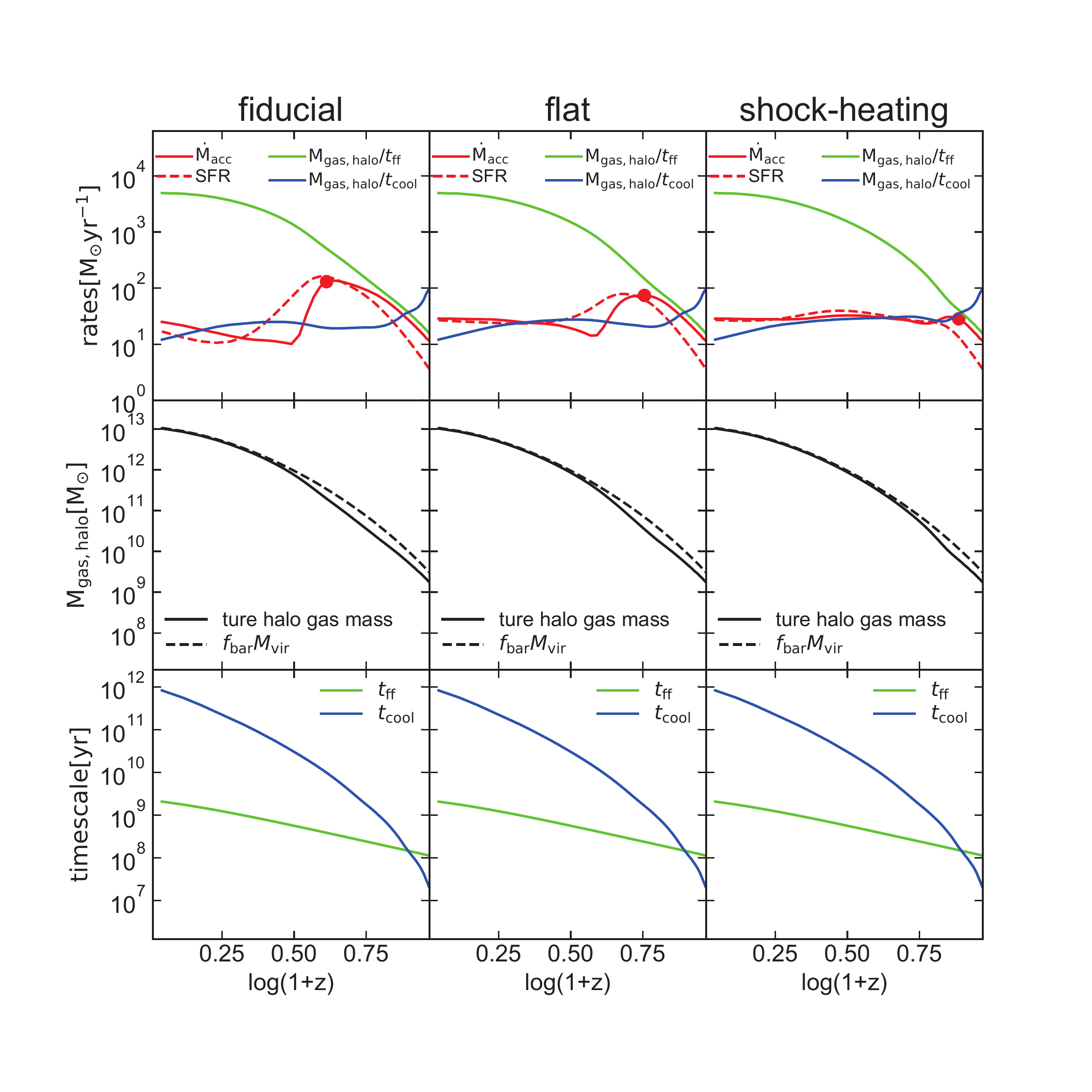}
    \caption{
	    Evolution of gas accretion and star formation rates (upper panels), the mass of halo gas 
	    (middle panels), and the free-fall and radiative cooling times (lower panels) 
	    for the most massive halo with $M_{\rm vir}=10^{14}{\rm M}_\odot$.
	    Red solid and dashed lines in upper panels respectively show the accretion rates and SFR 
	    in the model in which only gas accretion 
	    is included. 
	    Circles on red solid lines 
	    indicate epochs when the halo enters the cooling flow regime (i.e., hot-mode accretion).
	    Solid and dashed lines in middle panels are respectively 
	    the mass of the halo gas and the nominal 
	    mass $f_{\rm bar}M_{\rm vir}$, where $f_{\rm bar}=0.17$ is the adopted cosmic baryon fraction.
	 Because the gas accretion
	continues to consume the halo gas reservoir, the actual halo mass is  smaller than 
	the latter, although their difference is small. 
	    Bottom panels, the same for all accretion schemes, indicate the radiative cooling time 
	    and the free-fall time.
	    Green and blue lines in upper panels plot 
	    the  actual halo gas mass divided by the free-fall and radiative
	    cooling times, respectively.
    }
\end{figure*}

\begin{figure}
	\includegraphics[width=1.0\linewidth]{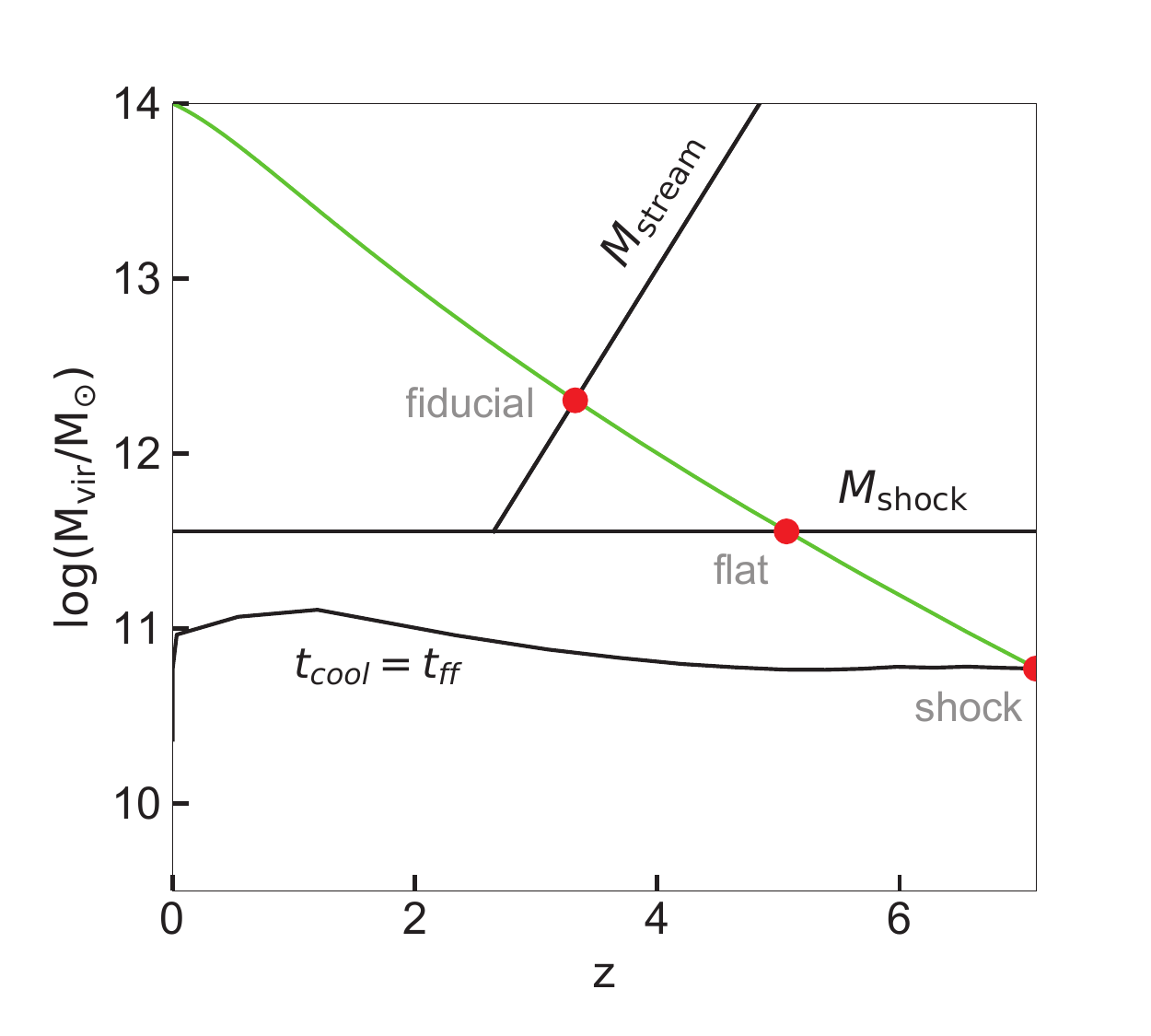}
    \caption{
	    Comparison of transition from free-fall accretion to cooling-limited accretion
	    in three different accretion schemes. The green curve indicates the trajectory 
	    of a halo with the present virial mass of $10^{14}{\rm M}_\odot$.
	    Red circles denote the transition for the scheme indicated.
    }
\end{figure}

\section{Gas accretion schemes and model families}

The first important question we must address is how the characteristics of gas accretion to discs
influence the star forming activity in galaxies. To tackle this problem, we construct three 
kinds of model families employing different gas accretion schemes.

\subsection{The fiducial scheme}

The fiducial scheme assumes the existence of cold-mode accretion as revealed in recent cosmological 
simulations. 
Cosmological simulations and analytical consideration
 predict three different regimes for the properties
 of the gas distributed in the dark matter halos depending on
 the virial mass and redshift 
      \citep[e.g.][]{de06,oc08}.
 They are demarcated by  two characteristic mass scales as shown in Figure 1 (left panel).
    One is the shock mass, 
      $M_{\rm shock}$, above which the halo gas develops a stable shock that heats the gas
      nearly to the virial temperature and the other is the stream mass, 
      $M_{\rm stream}$, below which part of the halo gas remains cold and is confined 
      into narrow filaments that thread the smoothly distributed shock-heated gas.
      The latter mass scale is valid only for high redshifts.

      The halo gas in Domain F is unheated and expected to accrete in free-fall to the inner region (the disc plane).
      In domain G, the gas heated to the virial temperature 
      attains near hydrostatic equilibrium
      in the halo gravitational field
      and the radiative cooling induces cooling flow to the centre
      with the cooling timescale.
      The gas behaviour is not so clear in Domain H, where cold gas streams coexist with 
      the surrounding shock-heated hot gas. 
      They may behave independently and accrete with their own timescales 
      or they may interact with each other leading to modification 
      of accretion timescales. Because no detailed information is available, 
      we assume that the cold and hot gases in Domain H accrete
      with the free-fall time and the radiative cooling time, respectively.

It is difficult to pin down the transition borders, namely $M_{\rm shock}$ and $M_{\rm stream}$, 
which depend on various physical quantities 
of accreting gas 
with the gas metallicity being the most influential parameter.        
Cosmological simulations using different gas dynamical codes (e.g., SPH and AMR) 
show different results, with the Eulerian mesh codes (e.g., AMR) tending to suppress
the cold fraction compared with the Lagrangian codes (e.g., SPH) \citep[e.g.][]{ne13}.
We use the average of the results from several studies summarized in Figure 6 of \citet{oc08},
keeping in mind that the adopted prescription may suffer from ambiguity and affect 
quantitative detail of the result.
Trial with several variations for the transition borders, however, revealed that 
the qualitative nature of model evolution is conserved within plausible parameter ranges.

\subsection{The flat scheme}

Existence of Domain H is a distinguished characteristic of the cold mode accretion. In the previous work
\citep{no20}, it was related to the formation of galactic bulges.
 The transition from the cold to hot modes corresponds to the quenching of star formation because 
 the gas accretion slows down abruptly at this point.
 The reconstruction of galaxy assembly and SFH 
 for the redshift range $z =1-10 $ by \citet{be19} indicates  
  increase of the quenching mass with redshift, though less steeply than 
 the theoretical prediction of \citet{de06}. 
To elucidate the effect of Domain H on the galactic SFH, we run an artificial model family in which this domain is removed
(middle panel in Figure 1).

\subsection{The shock-heating scheme}

The shock heating picture has been routinely assumed in many theoretical studies of galaxy formation.
It is interesting to see the difference between the star formation history derived under the fiducial
scheme and this picture.
There are two domains in this scheme separated by the equality 
of the radiative cooling time and the free-fall time (Figure 1, right panel).
When the gas cools with timescale shorter than the free-fall time, it is made to
accrete with the free-fall time. Otherwise, the gas is assumed to accrete with the cooling timescale.

\vspace{5mm}

For each accretion scheme, we run a series of galaxy models with the present halo virial masses varied in the range
$10^{10.36}{\rm M}_\odot \leq M_{\rm vir,0} \leq 10^{14}{\rm M}_\odot$.
This upper mass limit roughly corresponds to the halo mass inferred for
the most massive (disc) galaxies with the stellar mass $\sim 10^{11.5} {\rm M}_\odot$ 
on the basis of the abundance matching analysis \citep[e.g.][]{mo13,ro15,be19}. 
The lower mass limit 
corresponds to the stellar mass of 
$\sim 10^{7} {\rm M}_\odot$, below which the observational material becomes sparse.
Grey lines in Figure 1 show the evolution of each halo, illustrating how it goes through  
different regimes of gas accretion as it grows in mass.
Because no galaxy merger is considered in this study, the present models are regarded as mainly describing 
the evolution of galaxies whose growth rates are inflow dominated.

It is remarked here that the schemes illustrated in Figure 1 may need some modification.
As seen later, preventive feedback which acts to reduce gas accretion to the galactic disc,
is found to be important especially for low mass halos. The red line in each panel of Figure 1 indicates
the critical halo mass assumed in this study
for which this effect starts to be active.
Although the specific mechanism responsible for this process is not clear, previous studies suggest that
it affects a wider range of mass than the ionization by background radiative fields.
For comparison, we also plot the 'filtering mass' by \citet{gn00} and the 90$\%$ suppression
line by \citet{ok08} as a typical estimate for the radiative effect.
Preventive feedback, either on to discs or on to halos,
may thus significantly alter the picture solely based on the
cooling argument (and assumption of a universal cosmic baryon fraction).
We revisit this issue later.

\section{Comparison between different accretion schemes}

We first examine and compare the behaviours of the models for the three different 
gas accretion schemes.
We also compare the result with available observational data. 
 In this section, the parameters describing ejective feedback, preventive feedback, and gas recycling 
 are set to default values as given in Appendix.

\subsection{Star formation history}

Figure 2 shows the stellar-to-halo mass ratio as a function of the halo virial mass
for different redshifts. The halo mass here denotes the mass at each redshift.
The empirical analysis by \citet{be19} using the abundance matching method shows no significant 
difference  between the star forming galaxies and all galaxies including quiescent galaxies.
At the present epoch ($z \sim 0$), all schemes give similar results approximately agreeing 
with the result by \citet{be19} and other observational studies. Moving to higher redshifts ($z>2$), 
the shock-heating scheme progressively underproduces stars 
especially for more massive halos.
This is due to earlier transition of massive halos to the hot-mode gas accretion and 
associated weakening of star formation.
The predicted redshift dependence seems to be significantly stronger than 
the observationally inferred one even if we make allowance for the possible systematic deviation 
as suggested by the small ($\sim 0.3$ dex) overproduction 
in the low mass domain ($M_{\rm vir} < 10^{12} {\rm M}_\odot$) common to all three schemes.

Figure 3 shows the star formation history (SFH) for different values 
of the present halo masses. 
It is seen that the fiducial scheme captures several key features of the observed SFHs.
Namely, the SFR in halos with larger present masses attains the peak at earlier epochs and
declines more drastically after the peak. The halos in the lowest mass bin (the blue line) 
show monotonically increasing SFR. 
Notable dicrepancy is seen at z>2 where the models underpredict SFR compared with the observation.
This may indicate that the adopted feedback is too strong. However,
the discrepancy for most massive halos appears also in absence of feedback as seen later in section 5.
We discuss possible reasons for this discrepancy there.

The flat scheme is unable to reproduce this mass dependent behaviour. This indicates that the existence
of the domain H defined by the existence of cold gas streams contributes much in producing 
the observed mass dependent SFH, especially the early drastic 'quenching' in massive halos. 
The shock-heating scheme also fails due to the same reason.
It is interesting that \citet{be19} compare their results for the quenching fraction (their Figure 30)
to the prediction of \citet{de06} cold accretion model and find rough agreement, i.e., 
the domain for star forming galaxies extends to higher halo masses at higher redshifts.
This is in line with the present result showing the importance of the cold-mode accretion
of gas filaments penetrating through the surrounding shock-heated halo gas
although some other mass-dependent process (e.g. mergers) might bring about 
a similar effect.

	Differences in SFHs in the three schemes displayed in Figure 3 appear in high-mass domain with 
	${\rm log} M_{\rm vir}[{\rm M}_\odot]>12.5$, where accretion processes are different.
	Specifically, massive galaxies in the fiducial scheme experience peak SFR in earlier 
	epochs and larger drop in SFR after the peak than those in the flat scheme while 
	massive galaxies in the shock heating scheme show no peak.
	Figure 4 is meant to clarify the origin of these various SFHs going back to the difference 
	in the prescribed
	accretion recipes.
	Top panels show  the gas accretion 
	rates (red solid lines) and SFRs (red dotted lines) for 
	${\rm log} M_{\rm vir}[{\rm M}_\odot]=14$
	in the three accretion schemes.  For aid in analysis, middle panels show masses of the halo gas 
	and  bottom panels plot the radiative cooling time (blue) and the free-fall time (green) for the halo gas.
	We here turn off supernova feedbacks, gas recycling and 
	preventive feedback to explore more directly how the halo gas mass and the two timescales 
	eventually control the SFR under accretion-driven situation.

	Top panels indicate that SFRs essentially follow gas accretion rates with small time delay.
	Gas accretion rates in turn can be approximated by the amount of halo gas divided by the 
	accretion timescales. 
	Middle panels show that the  mass of halo gas (solid) is almost the same in 
	the three schemes. Therefore, the differences in SFHs should originate mostly in those in 
	accretion timescales.
	Green and blue lines in upper panels are 
	the 'accretion rates' 
	calculated by dividing the true halo gas mass respectively 
	by the free-fall time and radiative cooling times.
	These are again similar in the three schemes reflecting the similarity in the halo gas mass.

	The key for interpreting different SFHs lies in how the transition from the cold-mode accretion to the hot-mode accretion occurs.
	At the transition, the accretion timescale is switched from the 
	free-fall time to the radiative cooling time. 
	Figure 5 compares this transition for three accretion schemes. 
	In the shock-heating scheme, the transition occurs exactly when the 
	cooling time overshoots the free-fall time, but in the other two schemes it occurs when 
	the initial cold-accretion phase ends, which is later than the equality of two timescales.
	Instability of shocks in the flat scheme delays the transition compared with that in the shock-heating scheme.
	Existence of cold streams in high-mass halos further retards the transition in the fiducial scheme.
	Upper panels in Figure 4 overplot the transition epoch by red circles on the accretion rate curves.
	It is seen that the actual accretion rates (red 
	lines) follow closely the green lines before this transition epoch but they deviate and approach
	to blue lines after that.
	Comparison of three accretion schemes thus reveals that the difference in the accretion history
	comes from the difference in the transition epoch. 
	Moving from the shock-heating to flat to fiducial schemes, the transition epoch moves 
	progressively to more recent epochs and therefore the drop in accretion rate 
	gets larger because of larger difference in timescales at transition. 
	This explains why the peak SFR occurs latest and the decrement in SFR after the peak is
	the largest for the fiducial scheme as shown in Figure 3.

The SFHs such as plotted in Figure 3 will contain all information 
about the star formation activity in galaxies if accurately derived over the whole 
mass range. However, halo masses in the 
empirical approach are not obtained for each sample galaxy but derived statistically 
from abundance matching. On the other hand, it is not 
easy either to measure the halo masses for individual galaxies more directly, for example, from satellite 
kinematics \citep[e.g.][]{mo11} or 
weak gravitational lensing \citep[e.g.][]{ma06}. Any derived SFHs as a function of halo mass could thus be subject to uncertainty 
even if SFRs are obtained precisely.
We examine in the next section the dependence of star formation activity on the stellar mass, 
which is more directly assessed than the halo mass in general.

\begin{figure*}
	\includegraphics[width=1.0\linewidth]{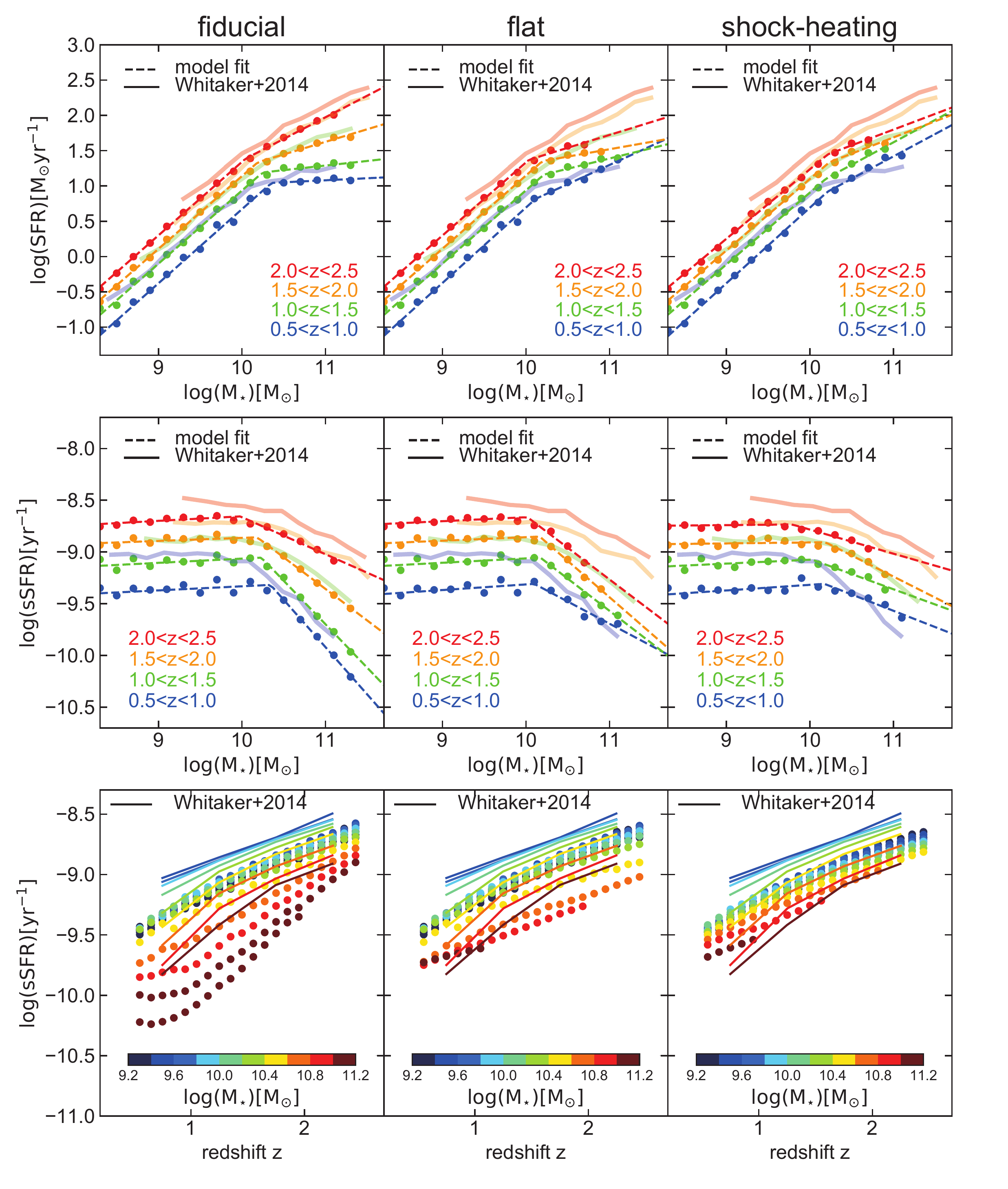}
    \caption{
	    Top panels: Star formation rate as a function of the stellar mass for 
	    different redshift bins. 
	    Filled circles indicate the model result with
	    the dashed lines showing the least squate fit of the model result by broken power law.
	    The solid broad lines indicate the result for star-forming galaxies by \citet{wh14}.
	    Middle panels: Same as the top panels but for the specific star formation rate.
	    Bottom panels: Specific star formation rate as a function of redshift for different stellar
	    mass bins. The solid lines are corresponding curves obtained by \citet{wh14} using 
	    UV+IR data.
         }
\end{figure*}

\begin{figure*}
	\includegraphics[width=1.0\linewidth]{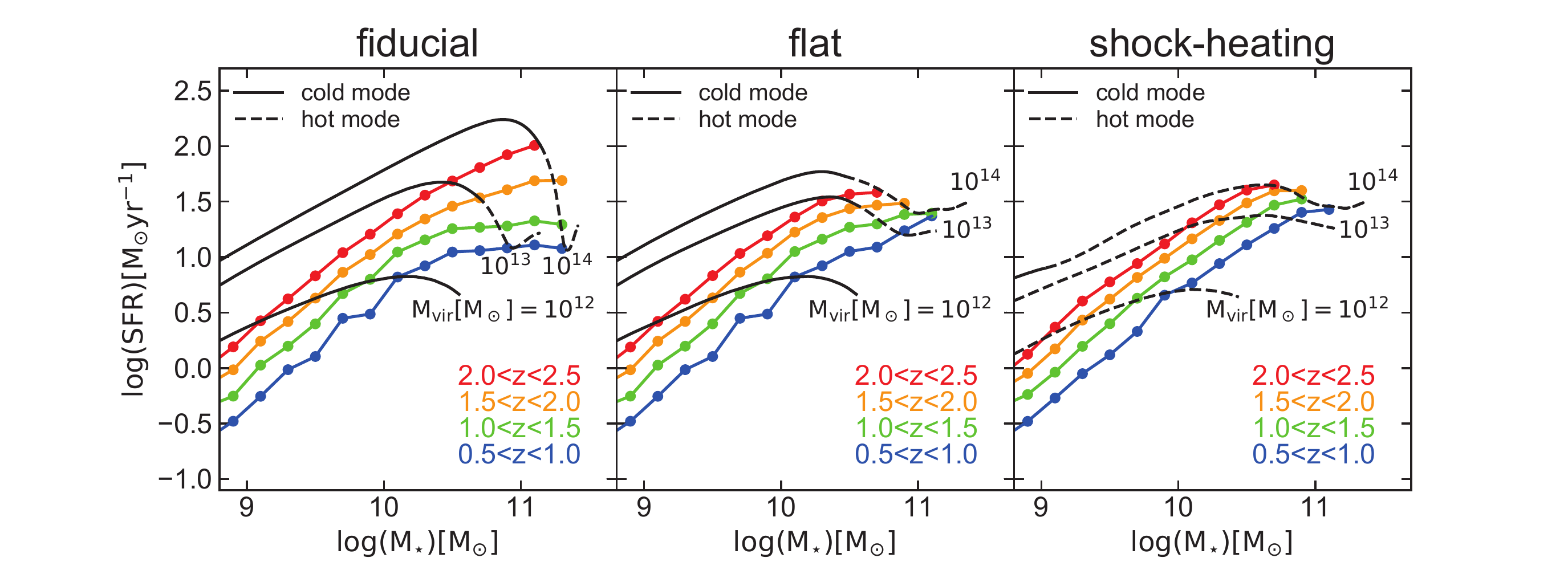}
    \caption{
	    Evolution trajectory of model galaxies overplotted on the model main seuqnece.
	    The main sequence is the same as the one plotted in Figure 6.
	    Trajectories are from top to bottom for the halos with ${\rm log} M_{\rm vir}=14,13,12$,
	    respectively. Solid and dashed parts denote the period of cold-mode and hot mode accretion,
	    respectively.
         }
\end{figure*}

\begin{figure*}
       \includegraphics[width=\linewidth]{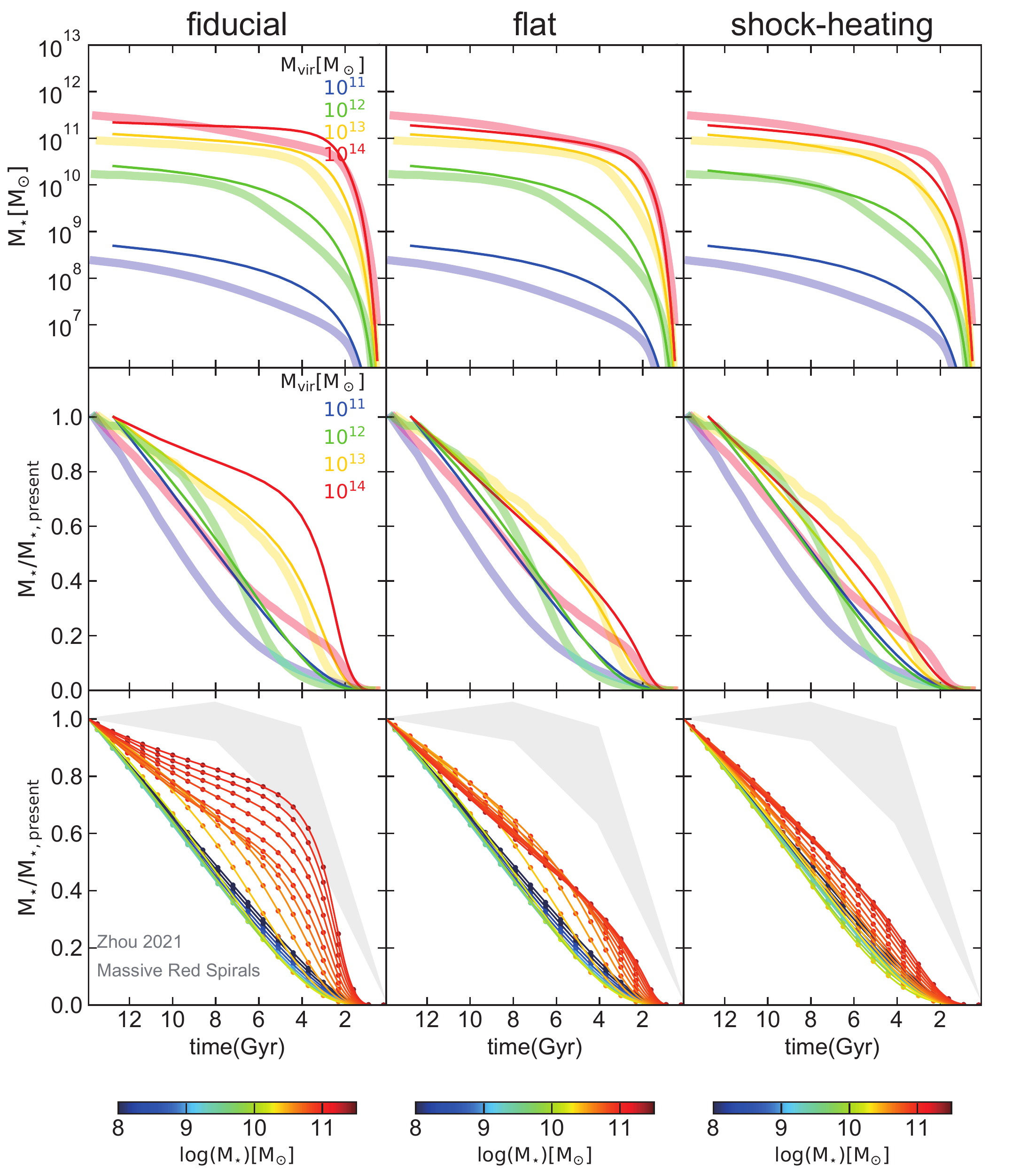}

    \caption{
	    Stellar mass growth history for different masses.
	    Top row: Thin solid lines indicate the time evolution of the stellar mass for  
	    different halo masses. Red, yellow, green, and blue indicate
	    the present virial masses of $10^{14},10^{13},10^{12}$ and $10^{11} {\rm M}_\odot$,
	    respectively. Broad shaded lines show the result by \citet{be19} for
	    all galaxies with the corresponding mass ranges.
	    Middle row: same as top row but plotting the normalized stellar mass against the 
	    cosmological time.
	    Bottom row:
	    Same as the middle row but as a function of the stellar mass at the present epoch.
	    The shaded region illustrates the inferred evolution for red massive spiral galaxies by \citet{zh21}
	     with the stellar masses $10^{10.5}-10^{11.5} {\rm M}_\odot$.
         }
\end{figure*}

\subsection{Main sequence evolution}

The correlations between star formation properties and galaxy stellar masses have 
been intensively examined in previous studies.
Among various scaling relations, 
the star formation main sequence (MS) of galaxies, i.e., the SFR-$M_\star$ relation,
provides  useful information on how 
galaxies with different masses formed stars over the cosmic time 
\citep[e.g.][]{wh14,jo15,sc15,ku16,to16,sc17,po19,co20,kh21} 
and serves as an independent
constraint on theoretical modelling \citep[e.g.][]{fe20}.

Figure 6 compares the SFR and sSFR predicted by the models with the observational data.
It is seen that the fiducial scheme reproduces several important features at least qualitatively.
Regarding the ${\rm SFR}-M_{\star}$ relation (top panels), it gives the SFR that increases almost linearly
with the stellar mass for the low-mass range, with normalization increasing with redshift.
The relation at higher masses bends in progressively larger degree for lower redshifts.
This behaviour agrees with the observational results about the redshift evolution of the 
normalization of the star formation MS and the turn-over in the high mass range \citep[e.g.][]{wh14,to16}.
Both the flat and shock-heating schemes give a reverse trend at high masses with smaller bending 
at lower redshifts.
The middle panels plot the sSFR as a function of the stellar mass. The fiducial scheme
shows nearly horizontal relation for low masses and increasingly steeper decline at high masses
toward lower redshifts, agreeing with the observational analysis.
The other two schemes fail to reproduce the observed trend.
Although we take here the observation by \citet{wh14} as a typical example, other observational studies 
also reveal similar trends \citep[e.g.][]{sc15,to16,ro20}.

Difference in the turn-over at high masses in different accretion schemes is directly related 
to difference in the SFHs. In order to clarify this point, Figure 7 overplots the trajectory
of evolving model galaxies on the ${\rm SFR}-M_\star$ relation. It is evident that large 
flattening at high mass end in the fiducial scheme is caused by strong drop in the SFR when 
the gas accretion is switched from the cold mode (solid curve) to the hot mode (dashed curve)
in massive halos with $M_{\rm vir}>10^{13}{\rm M}_\odot$. During the hot-mode phase, $M_\star$
increases by only 0.2 dex while SFR decreases  by almost 1 dex in the most massive halo.
The decrement in SFR is larger for more massive halos that move down on more right sides 
of the ${\rm SFR}-M_\star$ diagram, leading to 
the fan shape pattern of redshift sequence in the fiducial scheme.
High-mass end of the main-sequence is related to the hot mode accretion also 
in the flat and sock-heating schemes. In these cases, however, the decrease in SFR 
is milder and the stellar mass increases with almost constant SFR. Namely, the galaxy moves rightward 
nearly horizontally in the diagram. This means that the main sequence itself moves rightward.
Combination of this movement and the weak turn-over at high stellar masses at each redshift 
makes a configuration of the redshift sequence converging to higher stellar masses.

There has been much debate about the existence of the turn-over at high masses \citep[e.g.][]{sa17,po19}.
\citet{sp14} argue that the MS is nearly linear and its slope and normalization decrease
toward recent epochs, dismissing the existence of turn-over.
The slope at the high mass domain likely depends on the selection criteria for star-forming galaxies \citep{jo15}.
Galaxy morphology is also suggested to affect the high-mass behaviour of the MS. For example,
including  galaxies with dominant bulge components, usually characterized by smaller sSFR
compared with the bulgeless galaxies of similar masses, was found to create strongly 
downward bending at high masses 
as reported by \citet{ck20}. \citet{te16} also suggest close empirical relationship between the bulge mass 
(or the bulge-to-total mass ratio) and the star formation activity.
There is also a possibility that the star formation properties
and the main sequence characteristics depend on the 
galactic environments 
 \citep[e.g.][]{el07,gn11,ze13,ji18,nr18,na20,ol20}.
We return to these issues later in Discussion.

The bottom panels of Figure 6 show the redshift evolution of the sSFR for different stellar masses.
The fiducial scheme produces steeper decline of sSFR with time for more massive galaxies in qualitative agreement with data, 
whereas other two schemes show opposite trends in contrary to observations. 
This is a reflection of different 
turn-over behaviours seen in top panels. 
Absolute value of sSFR is underpredicted commonly in all accretion schemes
for ${\rm log M}_\star[{\rm M}_\odot]$ < 10.4. For 10.4<${\rm log M}_\star[{\rm M}_\odot]$<11.0, 
the fiducial scheme gives closer 
match to the observational data than other schemes. Finally, 
for 11.0<${\rm log M}_\star[{\rm M}_\odot]$, the fiducial scheme exhibits smaller sSFR than other 
schemes.

\begin{figure*}
	\includegraphics[width=0.9\linewidth]{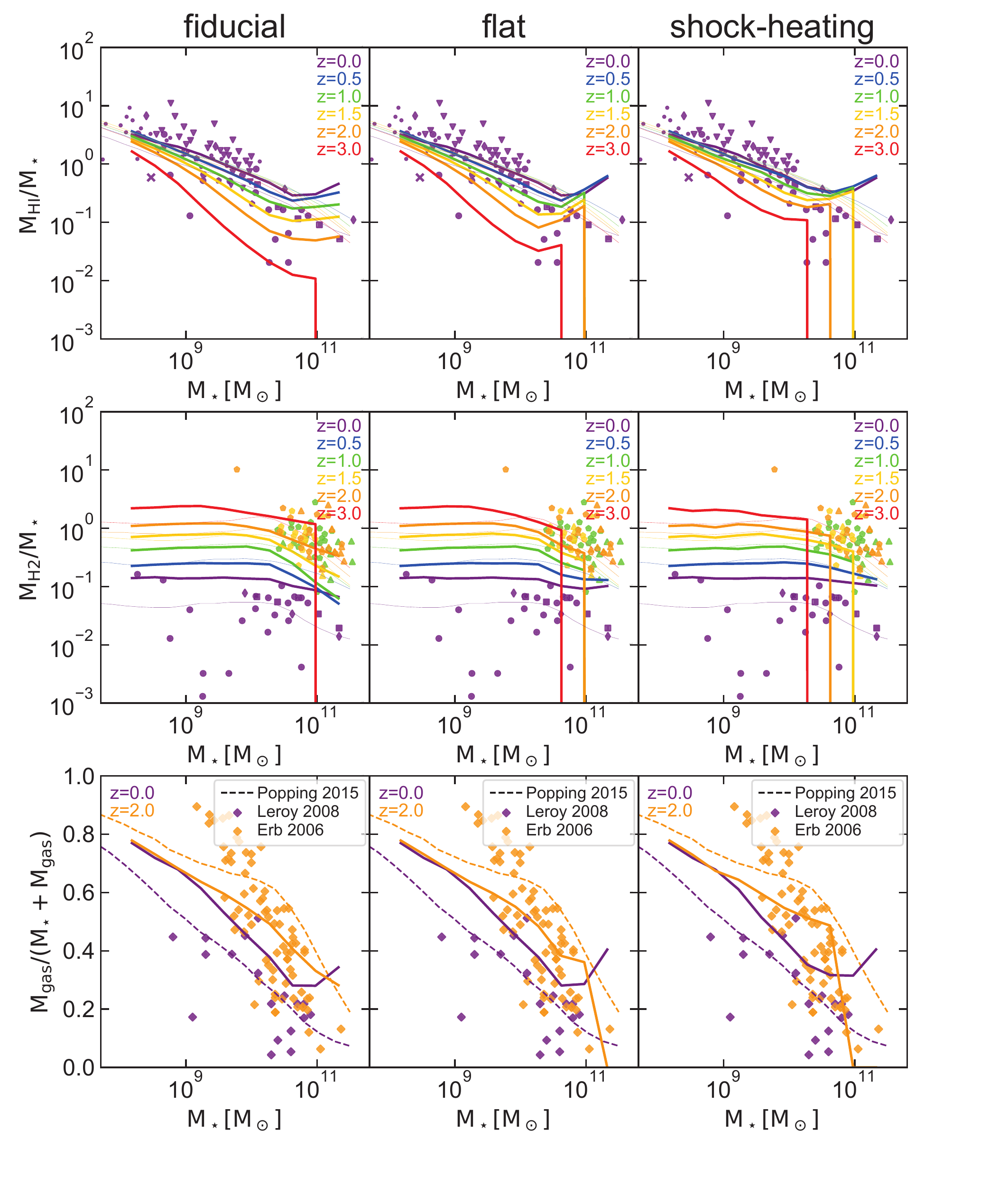}
    \caption{
	    The cold gas contents as a fiunction of stellar mass obtained in the present study
	    (thick solid lines) and the analysis by \citet{po15} (thin solid lines in the upper and middle 
	    panels and dashed lines in the bottom panels)
	    for different redshift bins 
	     with the width of 0.2 centered on the indicated redshift.
	    Top panels: The mass ratio of the atomic hydrogen and stars.
	    Observations from various literatures are indicated by symbols as follows.
	    squares: \citet{sa11}, diamonds: \citet{bo09}, circles: \citet{le08},
	    dots: \citet{mc12}, pentagons: \citet{bos14}, inverted triangles: \citet{bo14},
	    x: \citet{zh12}.
	    Middle panels: same as top panels
	    but for the molecular gas.
	    Observations from various literatures are indicated by symbols as follows. 
	    squares: \citet{sa11}, diamonds: \citet{bo09}, circles: \citet{le08},
	    triangles: \citet{ge10}, pentagons: \citet{ta13}.
	    Bottom panels: The mass fraction of the total gas (atomic+molecular) relative to the 
	    total baryonic (gas +stars) content as a function of the stellar mass
	    for $z=0$ (purple) and $z=2$ (orange). Observational values for $z=0$ are from direct 
	    measurements by \citet{le08}. Those for $z=2$ are the values from \citet{er06} 
	    who estimated the total gas mass from star formation rates.
         }
\end{figure*}

\begin{figure*}
	\includegraphics[width=\linewidth]{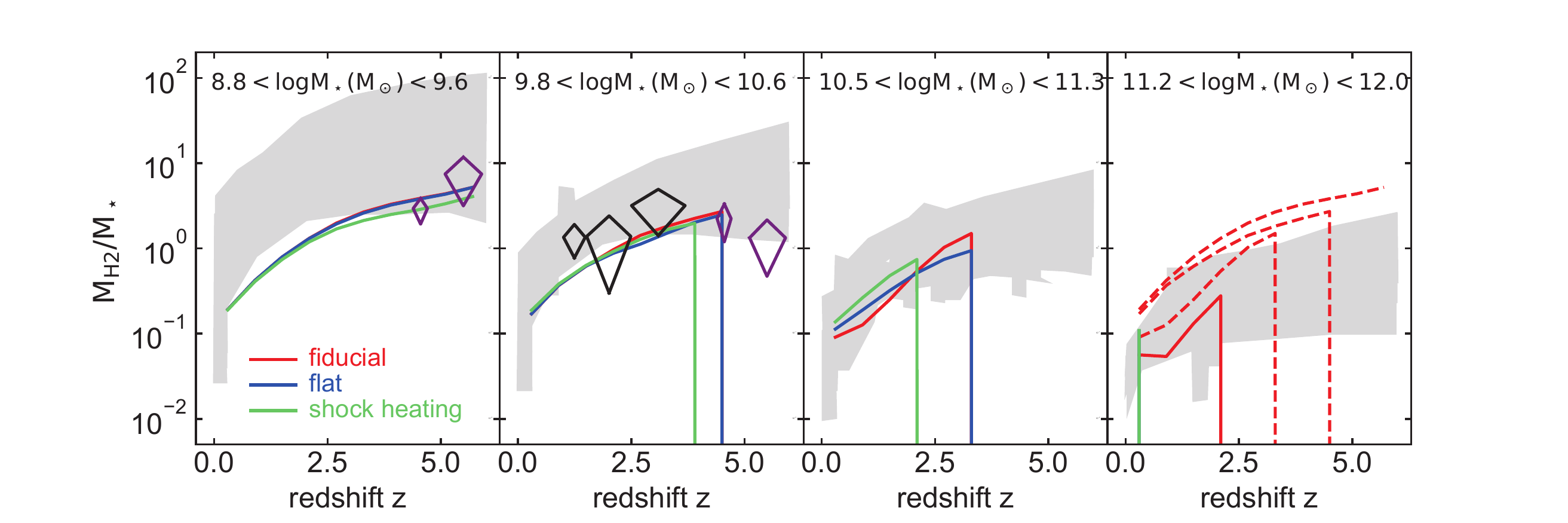}
    \caption{
	    Evolution of the mass ratio of the cold molecular gas and the stellar content
	    for different stellar mass bins in three accretion schemes.
	    Gray shaded regions summarise the observational data 
	    from the literature including \citet{sco17}, \citet{ta18}, and \citet{li19}.
	    Diamonds indicate the values and their errors
	    reported by \citet{de20} for the ALPINE [${\rm C}_{\rm II}$]-detected nonmerger galaxies (purple)
	    and the CO-detected galaxies from their compilation (black), respectively.
	    Three dashed lines in the rightmost panel repeat the less massive 
	    models in the fiducial scheme shown in left panels to illustrate the mass dependence
	    of the molecular content.
         }
\end{figure*}

The present result highlights the importance of gas accretion process in determining the shape 
and evolution of the star formation MS.
Especially, the shock-heating scheme seems 
to be incapable of explaining the qualitative behaviour of turn-over in the
 star formation main sequence claimed by several observational works.
This result agrees with that from several theoretical studies.
The semi-analytic models adopting the shock-heating paradigm for the halo gas 
\citep[e.g.][]{mi14,br15,he15,so15,wh15,gu16} tend to produce the sSFR-$M_\star$ relation 
that stays almost flat or even rises more steeply with the stellar mass toward more recent epoch,
being completely opposite to the observation.
The disc galaxy evolution models by \citet{du10} and \citet{la20} assuming the shock heating also produce the SFR-$M_\star$ relation
similar to these models.
On the other hand, the model by \citet{bo10} taking into account the cold-mode gas accretion produces
the SFR-$M_\star$ relation
that evolves only in the the normalization.
Cosmological simulations \citep[e.g.][]{ho14,vo14,fu15,sp15,sy15,do19,sc21} tend to
produce turn-over at high masses and show agreement with the observation 
to some extent but its reason remains unclear. One possibility is that these simulations,
treating gas processes more directly, automatically contain the cold-mode phase of gas accretion.

It should be noted that the main sequence turn-over may arise from 
other reason than specific feature of gas accretion. Generally,
any feedback process operating preferentially on massive halos  
would produce a qualitatively similar effect.
For example, 
the feedback from AGNs (when included) may contribute to the main sequence turn-over 
as argued in 
\citep{do19}.
In this specific case, AGN feedback should closely follow star formation. Quasar-mode feedback may be 
thus promising, though the coupling of feedback energy and the halo gas should be clarified 
to confirm its feasibility.

We focused so far on the star formation activity represented by the SFR and the sSFR.
We here examine the growth of stellar mass for different halo masses.
Figure 8 illustrates the stellar mass build-up in terms of the absolute mass (top row) and 
the fractional mass (middle and bottom rows). It is clear that the fiducial scheme explains the so-called
downsizing that more massive galaxies grow faster. Indeed, downsizing behaviour 
is already hinted in SFHs shown in Figure 3. This trend is more clearly shown 
by the fractional mass growth plotted in the middle and bottom rows.
In the flat and shock-heating schemes, the galaxies with different masses 
grow almost in concert. The grey shading in the bottom panels indicates the observed range 
derived by \citet{zh21} for massive red spirals. 
The fiducial scheme gives better reproduction than the other two schemes 
for the high-mass range. The reason is that the prolonged period of cold-accretion 
in this scheme allows more halo gas to accrete to the disc at early times (z>2)
and raises SFR above those in other two schemes as illustrated in Figure 4.
 It nevertheless  fails in matching the observation for $t>8$ Gyr.
This arises from the upturn in SFR for massive halos at z<0.5 seen in Figure 3 because it brings about 
significant stellar mass growth in recent epochs and makes early quenching impossible. This discrepancy 
could be 
erased by including, for example, AGN radio-mode feedback which likely operates for low gas accretion rate
in the hot-mode phase.

\subsection{Evolution of cold gas contents}

	Star formation rates in galaxies are controlled by the amount of cold gas that serves 
	as raw material for star formation and the timescale (depletion time) 
	for gas consumption. The amount of gas in turn could be determined 
	by several factors including the rate of gas accretion, ejection or removal rate of ISM, and 
	recycling efficiency of ejected ISM. 
Confrontation of the model prediction for the evolution of gas contents with  observational data
therefore helps judge importance of each factor in galactic star formation.
Gaseous content of the
cold interstellar medium is classified into atomic hydrogens (HI) and molecular hydrogens (${\rm H}_2$).
While the molecular gas is probed to high redshifts up to  $z=2 \sim 3$ thanks to the detection of CO lines,
 the observation of atomic component is
currently restricted to the local Universe, $z < \sim 0.2$, due to the bandshift effect of 21 cm emission lines.
Recent large surveys (e.g., BIMA SONG, HERACLES, COLD GASS) have revealed that the molecular 
masses relative to the stellar masses generally increase toward earlier cosmological epochs 
\citep[e.g.][]{he03,le08,ge10,sa11,ta13}.
Census for the atomic gas in the local Universe (e.g., ALFALFA, THINGS, GASS) 
consistently indicates a decreasing HI mass fraction toward increasing stellar masses
\citep[e.g.][]{gi05,wa08,ca10}.

Figure 9 compares the model result (thick solid lines) with the currently available observational data
(symbols). Redshift is colour-coded as indicated. The top panels show the mass ratio of the atomic hydrogen 
and the stellar content as a function of the stellar mass.
At $z=0$, all the schemes
give similar trends in broad agreement (a factor of $\sim$ 2) with the observation. 
As we move to higher redshifts, the difference between the three schemes becomes clearer, with the fiducial
scheme showing the largest decrease. The thin lines indicate the empirical result obtained by \citet{po15},
which shows very little evolution with redshift.
The simulation by \citet{da13} also suggests little evolution in HI content.
It is noted that the result by \citet{po15} was obtained indirectly from 
the observed SFR by using the abundance matching technique
combined with the locally established star formation law and the result by \citet{da13} is also based on
several assumptions in calculating HI fraction.
	If nevertheless these theoretical studies turn out to give better match to future observations,
	the shock-heating scheme might seem the most promising.
	There is a caveat here, however.
	Stronger decrease in HI content to higher redshifts for the fiducial scheme 
	than in other schemes may 
	have partly resulted from 
	more active cold accretion that produced discs with richer gas content. 
	 Gas richness (i.e., higher surface densities) in discs 
	means a relatively higher fraction of ${\rm H}_2$ in the disc and therefore poorer HI
	content. Another possible factor influencing gas contents is their spatial distribution.
	 In actual galaxies, ${\rm H}_2$ and HI are spatially decoupled, with 
	the former more concentrated to galactic centers \citep[e.g.][]{yo91}. The model assumes that 
	two components have the same scalelength  corresponding to the 
	halo spin parameter $\lambda$ of 0.06, which gives disc radii more relevant to the molecular 
	component. Increasing $\lambda$ to 0.08 as a compromise between two components 
	significantly lifts up HI content 
	relatively to ${\rm H}_2$ due to resulting lower gas densities. It also 
	diminishes difference in redshift dependence between 
	different schemes. weakening apparent advantage of the shock-heating scheme.
	Argument relying on the spin parameter is admittedly rough but this result suggests 
	a promissing avenue to resolve the difficulty shown here.

Middle panels of Figure 9 show the mass ratio of the molecular hydrogen.
Compared with HI, all the schemes show much weaker mass-dependence with only the normalization 
increasing with redshift. Weak mass dependence agrees with the observed trend. 
The observational data and the empirical model by \citet{po15} (thin lines)
indicate the mass ratio at $z=2-3$ larger than the local value 
by more than $\sim 1.5$ dex. All schemes give good match to this result
 but tend to overpredict
	$M_{\rm H2}/M_\star$ up to $\sim 1$ dex at $z=0$.
This likely  originates in  overabundance in 
 ISM gas  in model galaxies (especially massive ones) at recent 
epoch. Increasing gas ejection artificially by 1 dex for z<0.5  leads to lower molecular content and
better agreement  at z=0. This modification also helps erase turn-up 
of model SFRs at $z<0.5$ in massive halos seen in Figure 3 for the fiducial scheme. 
and improve agreement with the observation. Although we boosted here the ejection rate by supernova 
feedback artificially, more promising alternative mechanism to operate is radio-mode AGN feedback that
likely acts preferentially for massive halos (Domain G of the fiducial scheme) at low redshifts.

Bottom panels of Figure 9 show the mass fraction of the total gas including atomic and molecular hydrogen.
The three schemes all show an increasing fraction toward smaller stellar masses, with the value at  
fixed mass at $z=2$ larger than in the local Universe. These trends agree qualitatively with the observation
but the latter suggests larger ratios and steeper mass dependence at $z=2$ whereas the ratios
are smaller at $z=0$.
The dashed lines  indicate the result of the analysis performed by \citet{po15}.
Their result shows larger redshift dependence than our models, though the slope of mass dependence is similar.
 The discrepancy of the fiducial scheme and observation can be partly relieved by using combination of 
 the large spin parameter and boosting of gas ejection discussed above. Larger spins contribute 
 to boost HI content at $z=2$ without affecting the molecular content, leading to larger 
 gas-to-stellar ratio. Applied boosting reduces ${\rm H}_2$ content at $z=0$ reducing gas fraction.
 Two effect combine to produce larger redshift evolution in closer agreement with the observational data.
	 Exploring evolution of two gas components consistently requires knowledge of 
	density profiles of those comonents and reliable information about angular 
	 momentum distribution
	 of infalling gas, which control disc sizes.
	Radio mode AGN feedback is believed to operate in low efficiency accretion to massive 
	black holes provided by the hot-mode accretion \citep[e.g.]{fa12}. 
	Including growth of black holes is necessary to treat this process properly.
	We defer such investigation to future study.

Recent observations of the molecular gas can reach $z>5$.
Additional analysis was carried out of the evolution of the molecular gas over large 
redshift range. 
The result is shown in Figure 10 for different stellar mass bins.
This figure gives the essentially same information as the middle panels of Figure 10 but now shows the 
redshift dependence to earlier cosmological epochs.
The observations suggest the 
ratio of the molecular mass to the stellar mass increasing with redshift up to $z>5$ for all mass bins 
and larger ratios for lower mass galaxies 
at the fixed redshift. The three schemes considered here show a qualitatively similar behaviour to 
the observation although the discrepancy 
can reach $\sim 1$ dex at the lowest mass bin, where the three schemes give nearly identical evolution.
Although the fiducial scheme cannot reproduce every detail of the observational result, it
predicts strong evolution of the molecular content in agreement with the observation.
These analyses also suggest that the improvement in the observational data  for galactic gas contents 
(especially the redshift evolution of the atomic gas) is needed to 
 further constrain the models examined here.
The data expected to be obtained by the newest instruments such as
 Atacama Large Millimeter Array (ALMA) and
 Square Kilometre Array (SKA) will serve as a powerful constraint on theoretical models.

\section{Roles of feedback and gas recycling}

The discussion above highlights the potential importance of the cold-mode accretion especially for massive galaxies.
The cold-mode proceeding with free-fall time delivers halo gas to discs rapidly
supporting high SFR. In the fiducial scheme, this phase lasts longer than in other two cases
until the epoch when the radiative cooling time becomes far-longer (by $\sim1$ dex) than the free-fall time.
When the accretion is switched to hot cooling-limited mode, the accretion rate and therefore SFR
drops drastically by $\sim 1$ dex (see Figure 4),
 reproducing strong quenching observed in massive galaxies (Figure 3)
and inducing large main-sequence turn-over at high-mass end (Figure 7).

Here we examine the role of other physical processes (preventive feedback, ejective feedback,
and ISM gas recycling) which could modulate the evolution driven by gas accretion. 
For the reference model in this practice, we adopt a model in which only gas accretion of the fiducial scheme 
is included, which we call the accretion-only model.
We perform three kinds of modelling by adding ejective feedback, ejective feedback plus  
gas recycling, and preventive feedback separately to the accretion-only model.
By comparing each type of model with the accretion-only model, we aim to isolate the effect of each process 
added.
In the following, we restrict the gas accretion to the fiducial scheme and the parameters for each 
added process are default values unless otherwise stated.
We can reverse this procedure 
by removing each process  from the full model.
This was done and the conclusions about 
the respective role of those processes remain the same.

\begin{figure*}
	\includegraphics[width=\linewidth]{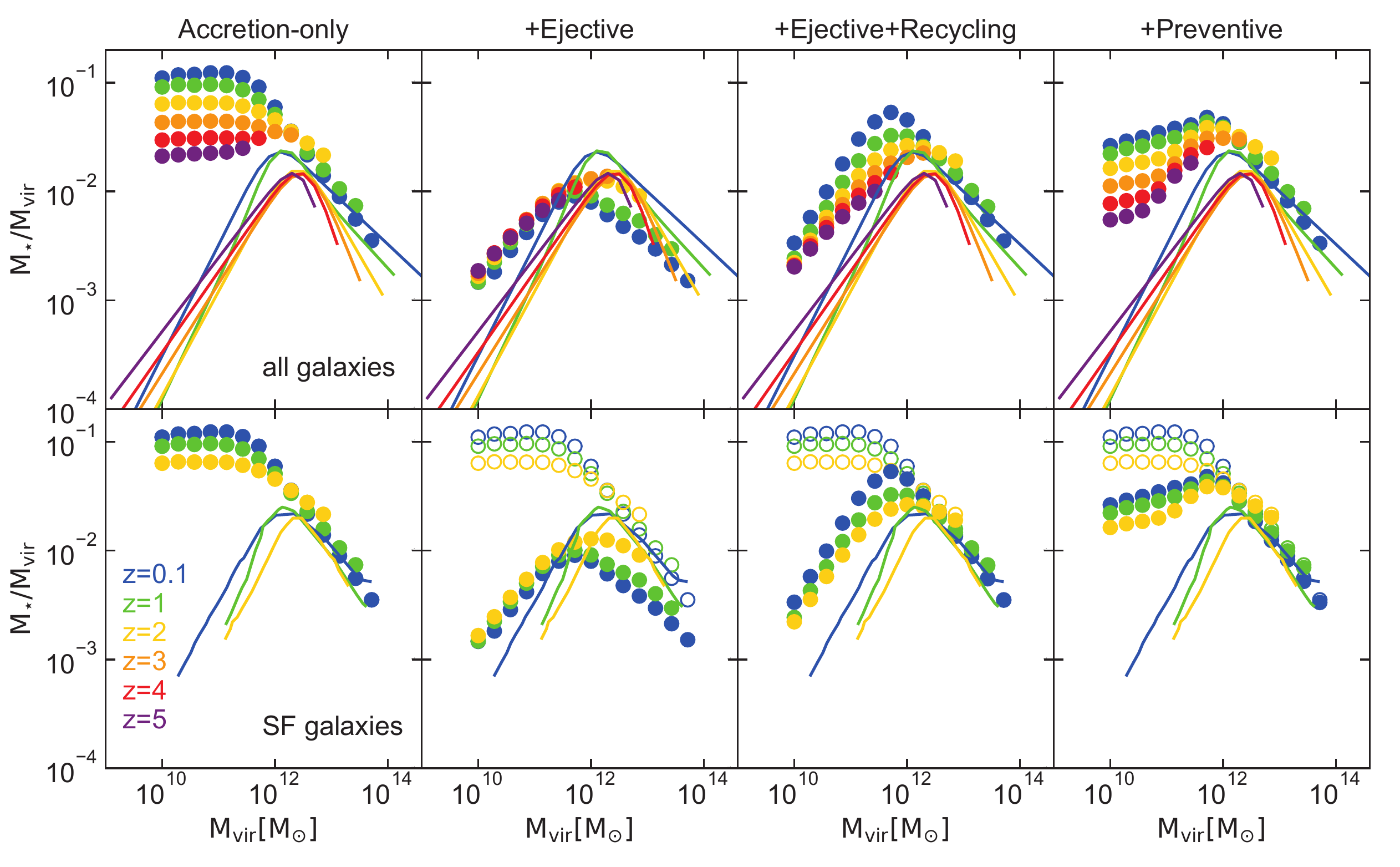}
    \caption{
	    Same as Figure 2 but depicting the effect of feedback and recycling. 
	    Circles indicate the models with redshift colour-coded as indicated.
	    The left-most columns show the result for the reference model in which only gas accretion
	    in the fiducial scheme is included. Second to fourth columns show from left to right
	    the models with
	     ejective feedback, ejective feedback plus recycling , and preventive
	    feedback added to the accretion-only model, respectively.
	    The accretion-only model is replotted in the right three columns with open circles 
	    for comparison (only lower panels).
	    The results obtained by \citet{be19} are plotted by solid lines in top panels (all galaxies) and 
	    bottom panels (star forming galaxies) for comparison.
         }
\end{figure*}

Figure 11 is similar to Figure 2 but indicates the effect of each process on the stellar-to-halo 
mass ratio. It is seen that the accretion-only model already reproduces the decrease in
this ratio toward higher halo masses consistent with the observation. On the other hand, for the low-mass range
($M_{\rm vir} < 10^{12} {\rm M}_\odot$), it gives a nearly constant mass ratio at each redshift,
overproducing the stellar content up to two orders of magnitude. Inclusion of ejective feedback 
leads to a profound reduction of the stellar mass especially for low mass halos, bringing their mass ratio 
close to the observation. However, stellar masses for massive halos fall below the observation. The 
net effect of gas recycling is to lift up the final stellar mass over 
the entire mass range by returning some portion of the ejected gas to the disc for 
additional star formation. This change enlarges the discrepancy for low halo masses but
remedies the deficit of 
stellar mass caused by ejective feedback in high-mass halos almost completely.
Preventive feedback by itself affects mainly the low mass halos with 
$M_{\rm vir} < 10^{12} {\rm M}_\odot$ and reduces their stellar masses by $\sim$ 1 dex
 with respect to the accretion-only model.

\begin{figure*}
	\includegraphics[width=\linewidth]{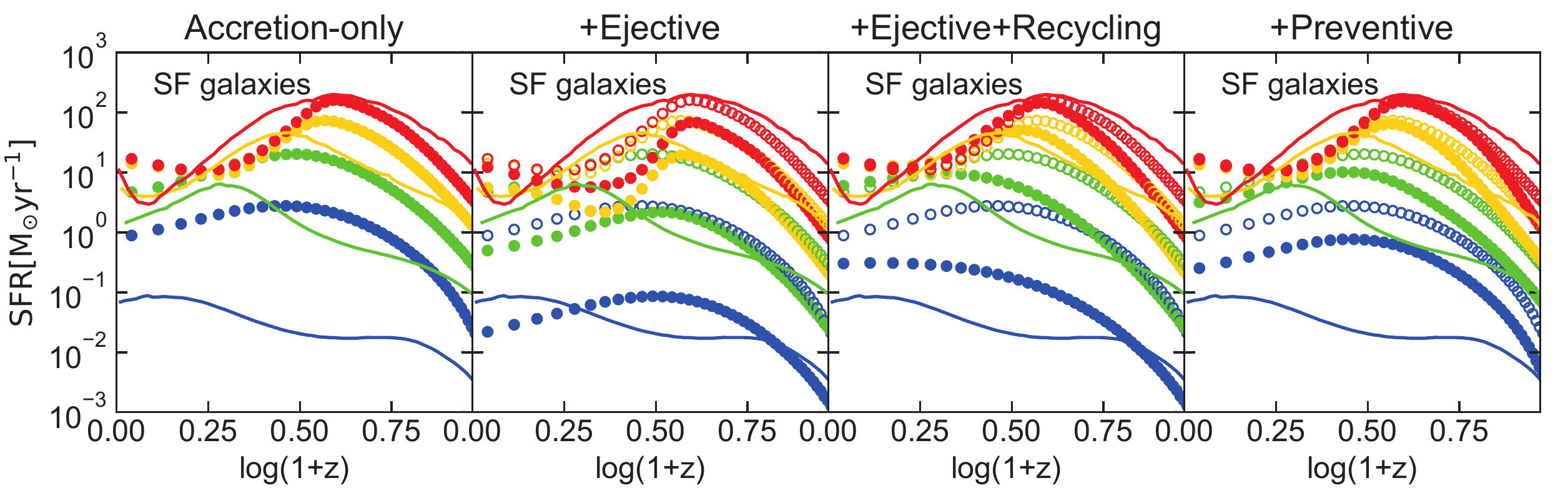}
    \caption{
	    Same as Figure 3 but depicting the effect of feedback and recycling.
	    Colour coding is the same as in Figure 3. Circles indicate the model 
	    while solid lines show the result by \citet{be19}.
	    The SFHs in the accretion-only model (left-most columns) are repeatedly shown 
	    by open circles in other columns for comaprison.
         }
\end{figure*}

Figure 12 is similar to Figure 3 but illustrates the effect of each process on the star formation
history.
It is noted that the accretion-only model already creates several important features 
in the observed SFHs.
Namely, the epoch of peak star formation is earlier and the fall of the SFR after the peak is larger 
for more massive galaxies especially in the high-mass range.

It is also noted that this model underproduces stars in massive systems at early times.
Several possibilities can be considered for this discrepancy. The star formation law, i.e., 
the scaling of SFR with gas densities, adopted in the 
model, which is based on the local-galaxy statistics, may not apply to high-z galaxies. 
Or the initial mass function (IMF) may be top-heavy at early times because
of low metallicity in the ISM, leading to  overestimate of the observed SFR. The latter possibility can be 
explored in principle by incorporating a time-dependent IMF in the model and calculating the SFR indicators used 
in the Behroozi's analysis for direct comparison. Considering the crudeness of the model
(e.g., the simplified treatment of gas accretion), it is not considered 
feasible at the moment.

All other models share the same feature as the accretion-only model for massive halos with 
$M_{\rm vir} > 10^{12.5} {\rm M}_\odot$, meaning that the cold-accretion
plays a critical role and the addition of other processes 
does not change the qualitative feature of the SFHs in massive halos appreciably.
Ejective feedback reduces the SFR during the whole cosmological epoch with its effect getting stronger 
for lower halo masses.
Gas recycling has a notable effect in low mass halos. It erases the peak in the SFR at $z\sim 2$ 
and changes the SFH into a monotonically increasing profile.
This qualitative change leads to improved match to the empirically inferred SFHs 
by \citet{be19}. This is due to the long timescale of the recycling for low mass halos 
implemented in the model. Namely, the recycling acts to delay star formation by taking a long time 
in returning the 
ejected gas to the disc. This change is recognized also for halos in the $10^{12} {\rm M}_\odot$ bin
(green) but is not strong enough to erase star formation peak completely.
Finally, preventive feedback helps suppress over-production of stars in low mass halos but does not 
affect the shape of the SFR profile (Compare with the accretion-only model).

\begin{figure}
	\includegraphics[width=\linewidth]{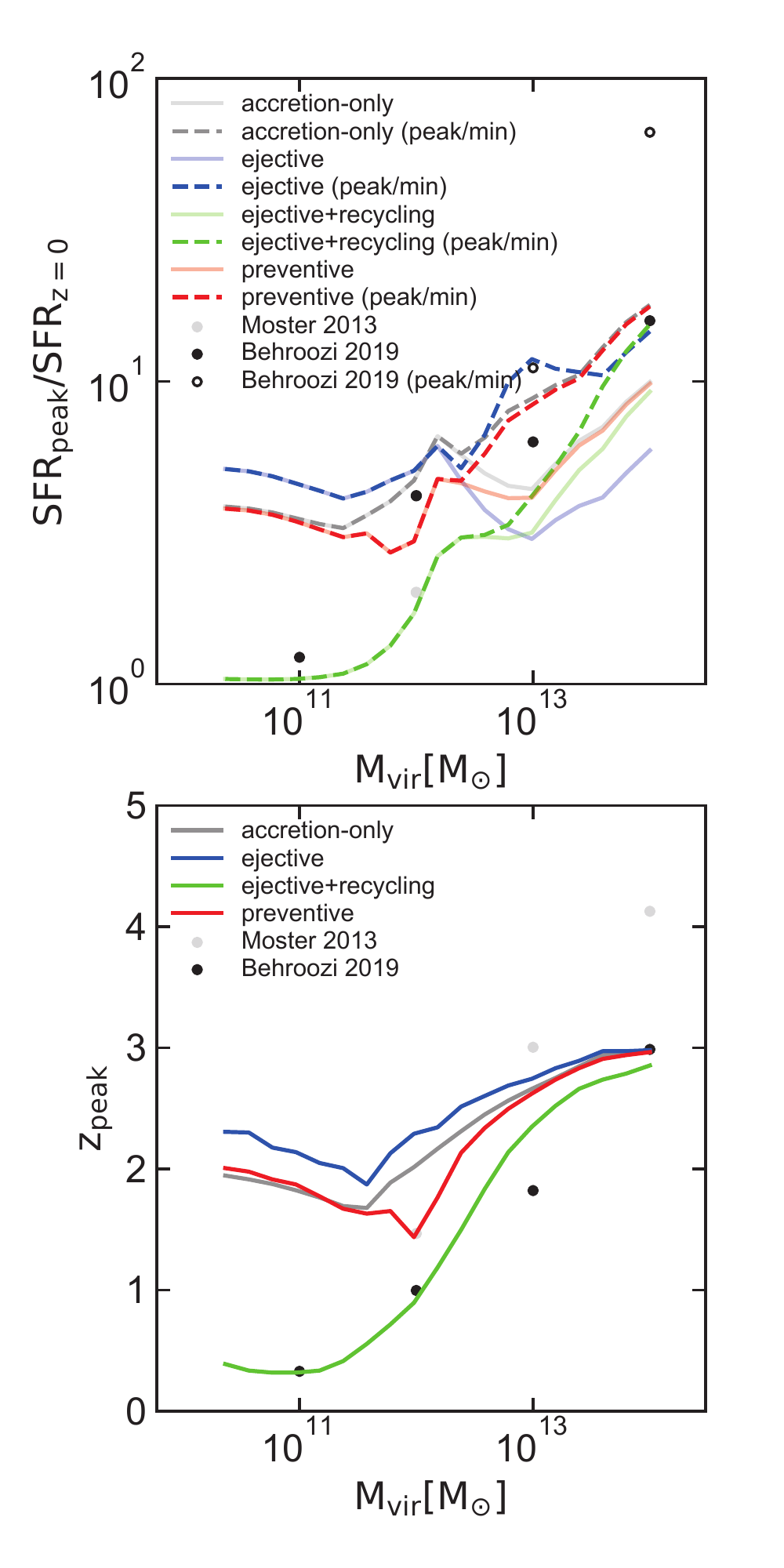}
    \caption{
	   Upper panel: The ratio of the peak SFR and the present SFR (solid lines)
           or the minimum SFR (dashed lines) as a function of the present halo virial mass.
           Lower panel: The redshift at which the peak SFR is attained as a function of the present halo virial
           mass.
           Grey and black circles show the empirically obtained values in \citet{mo13} and
           \citet{be19}, respectively.
         }
\end{figure}

The result mentioned here is summarized more quantitatively in Figure 13.
Ejective feedback and preventive feedback have negligible 
effect on the peak-to-minimum SFR ratio and redshift for the peak SFR, keeping the monotonically 
increasing trend with the halo mass. 
For massive halos, the peak-to-present SFR ratio are smaller than the peak-to-minimum one
due to later raise of the SFR toward present. This leads to much flatter mass-dependence of this quantity.
For small-mass halos, these two are equal because either the SFR monotonically decreases after the peak
 or the SFR monotonically increases with time.
The recycling process has a remarkable effect in low mass halos ($< \sim 10^{12} {\rm M}_\odot$),
moving the peak nearer to the present epoch.
This change is clearly indicated in the lower panel of Figure 13 that plots the peak redshift against the halo mass.
The accretion-only model shows the peak redshift taking the minimum at intermediate mass
and increasing on both sides. 
This feature is not
destroyed by either ejective or preventive feedback. The recycling brings the model in agreement with the 
observation over the whole mass range by greatly reducing the peak redshift in low mass halos. 
Recycling also reduces 
the peak-to-present SFR ratio almost to unity (upper panel). The peak epoch coincides with the present 
time for a monotonic SFR profile.

\begin{figure}
	\includegraphics[width=0.9\linewidth]{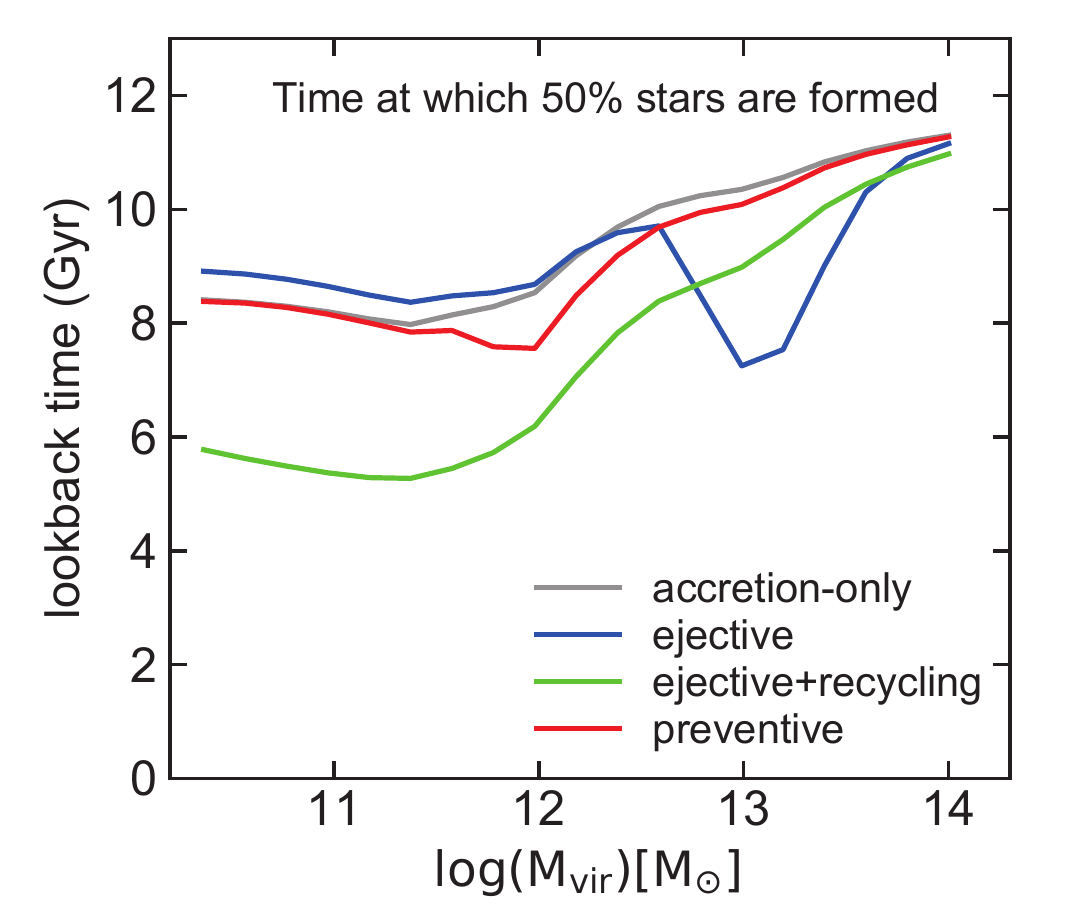}
    \caption{
	  The lookback time at which half the present stellar mass was formed as a function of 
	  the present halo mass. Different colours indicate models in which
	  different processes are included as indicated.
         }
\end{figure}

Figure 14 plots the lookback time at which 50\% of the present 
stellar mass is formed as a function of the present halo mass.
The accretion-only model shows this critical time 
increasing with 
the halo mass but the dependence is weak. Ejective and preventive feedback preserve this 
property though the halos around $10^{13} {\rm M}_\odot$ have their growth retarded considerably by
ejective feedback. Except this change, we can say that the model galaxies are on the whole fairly 
old, reflecting the dark matter halo growth history in a $\Lambda$CDM Universe. 
Gas recycling causes an appreciable change by slowing the stellar mass
growth in low mass halos down to $\sim 5$ Gyr ago, producing the downsizing behaviour. 
This change brings the model into a good agreement
with the observational inference and also agrees with the result by the SAMs
\citep[e.g.][]{he13,he15,mi15}. 
Delay in stellar mass growth for $10^{13} {\rm M}_\odot$ halos mentioned above 
is somewhat unexpected behaviour. We touch on this point later.

\begin{figure*}
	\includegraphics[width=\linewidth]{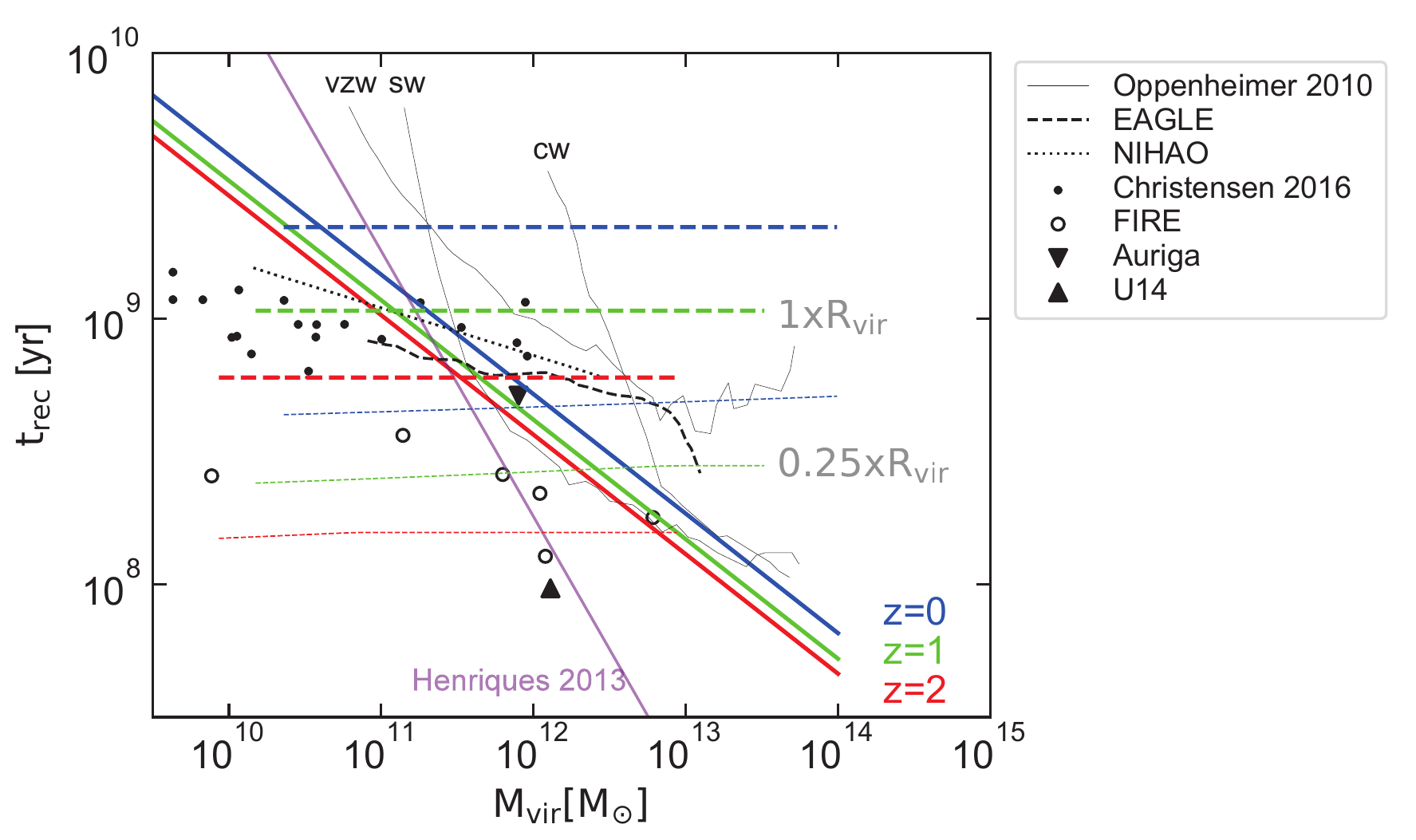}
    \caption{
	    The recycling timescle as a function of the halo virial mass and redshift.
	    Blue, green, and red lines indicate the dependence adopted in the present study.
	    The purple line indicates the redshift-independent relation proportional 
	    to $M_{\rm vir}^{-1}$ assumed in the semi-analytic model of \citet{he13} (their equation 8).
	    Others indicate the results of cosmological simulations for comparison.
	    Three black lines: three simulations with different feedback recipes by \citet{op10};
	    black dashed line: EAGLE simulation by \citet{mi20};
	    black dotted line: NIHAO simulation by \citet{to19};
	    black dots: the simulation by \citet{ch16};
	    open circles: FIRE simulation by \citet{an17}.
	    EAGLE, FIRE, and the simulation by \citet{ch16} give the recycling time over the range $0<z<3$.
	               Inverted triangle: Auriga simulation by \citet{gr19};
		                   triangle: the simulation by \citet{ub14}.
	     Two groups of coloured dashed lines indicate the free-fall time at 
	     the virial radius, $R_{\rm vir}$
	     (upper group of thick lines), and at $0.25 R_{\rm vir}$ (lower group of thin lines), respectively.
	    They also indicate the mass range for each redshift explored 
	    in the present model. 
         }
\end{figure*}
       
These results highlight the way in which  the qualitatively different SFHs in high-mass and low-mass 
halos are brought about by the dominant processes acting on each mass domain.
The transition from the cold mode to the hot mode of gas accretion makes a backbone for the SFHs in
massive halos with the present mass larger than $\sim 10^{12} {\rm M}_\odot$.
In low mass halos, gas recycling plays an essential role in delaying star formation
and changing the shape of SFH to a monotonically increasing profile.
This is not achieved by either ejective or preventive feedback.
 Inability of ejective feedback in shifting SF to recent epochs
and therefore making low-mass galaxies younger is also demonstrated by \citet{we12}. 
The delay seen here is due to the long recycling timescale for these halos.
We take a detailed look at this point here.

Figure 15 (solid coloured lines) shows the mass and redshift dependence of the recycling time scale $t_{\rm rec}$ assumed in the present study.
It is the prescription used in \citet{mi15}.
 The recycling timescale increases for lower halo masses 
and lower redshifts. 
For reference, the dashed coloured lines indicate the free-fall time at $R_{\rm vir}$ (upper group)
and $0.25 R_{\rm vir}$ (lower group), respectively. 
The former is representative of the halo free-fall time.
The latter radius defines roughly the inner radius of the circumgalactic region (CGR)
used in the analysis of cosmological simulations \citep[e.g.][]{mu15}.
The halo free-fall time depends on redshift
but not the halo mass due to the definition of the virial radius.
The free-fall time at $0.25 R_{\rm vir}$ is not strictly constant with the mass 
because of the mass dependence 
of the halo concentration. It should be also noted that the free-fall time here is calculated 
from the dark matter density profile and does not include either the contribution of the baryonic component or
the effect of the baryon-induced halo contraction. It therefore gives only the upper limit.

The free-fall time, $t_{\rm ff}$, decreases with decreasing radius. 
If $t_{\rm rec} < t_{\rm ff} (R_{\rm vir})$, there should be some radius $r_{\rm turn}$ where 
$t_{\rm rec} = t_{\rm ff} (r_{\rm turn})$.
If we assume that the ejected material reaches to the turning radius very rapidly and fall backs
almost ballistically (i.e., in free-fall fashion), then $r_{\rm turn}$ 
is considered to give
rough estimate for the maximum distance from the halo centre which can be reached by the ejected ISM.

It is seen that the adopted recycling time is  
comparable to or shorter than the halo free-fall time on the whole. 
Therefore, our prescription implicitly means that most of the ejected ISM cannot escape from the
parent halo into the intergalactic space. We note here in passing that the mass of the ejected ISM 
in our energy-driven feedback is calculated under the requirement that the ejecta have the escape velocity 
of the dark matter halo and can therefore reach to the virial radius unless decelerated by some mechanism(s).
The free-fall time at $0.25 R_{\rm vir}$ falls below 
the recycling time for the halo mass less than $\sim 10^{12} {\rm M}_\odot$ for all redshift.
The ejected ISM in these halos can thus reach the CGR.
On the other hand,
more massive galaxies have the recycling time comparable or shorter than the circumgalactic free-fall time 
$t_{\rm ff}(0.25 R_{\rm vir})$.
In this case, the ejected ISM must return to the disc very rapidly. Such a short recycling time 
physically allows the extreme case in which the feedback energy cannot push the surrounding ISM outside the 
disc but only puff it up. The notion of gas recycling loses its meaning in such a situation.
In this study, this occurs only in relatively massive halos 
with $M_{\rm vir}>10^{12} {\rm M}_\odot$, for which the recycling plays a subdominant role. 

Cosmological simulations give useful and more direct information for comparison.
It should be noted that the definition of the recycling time is not necessarily the same.
For example, \citet{ch16}, \citet{an17} and \citet{mi20} 
define $t_{\rm rec}$ as the time interval between ejection from the central galaxy and subsequent 
re-accretion on to its ISM. On the other hand, \citet{op10} and \citet{gr19} define $t_{\rm rec}$ 
 as the interval between ejection and re-ejection/
star particle conversion, which additionally includes the time spent by the ejected gas within the ISM as gas component.
\citet{to19} give only the time the ejected gas spends as cold circumgalactic gas (namely, that gas is not 
required to reaccrete on to the parent galaxy).
This situation makes the comparison of different simulations not straightforward 
(see discussion in \citet{mi20}). 

Bearing this caveat in mind, Figure 15 compares our recipe with these simulations.
The cosmological simulations by \citet{ch16}, EAGLE, and NIHAO indicate the recycling time of the order of 1 Gyr 
for halo masses smaller than $\sim 10^{12} {\rm M}_\odot$ that corresponds to 
$M_\star <  10^{10} {\rm M}_\odot$, in rough agreement with the prescription adopted here, 
 although the mass dependence suggested by these simulations is much weaker.
The simulations by \citet{op10} give systematically longer recycling times for most of the mass range
whereas FIRE simulation by \citet{an17} indicates shorter timescales.
Our prescription lies roughly in the domain spanned by these simulations.

As discussed above, the recycling time prescription
adopted in our study is not incompatible with the conceptual process that the part of the ISM  launched 
by feedback is delivered 
deeply into the circumgalactic space and re-accretes to the disc after a certain time.
In relation to this, cosmological simulations give direct measurements of the maximum distance the ejected gas travels before
falling back to inner regions.
\citet{gr19} report the typical maximum distance 
of $\sim$ 20 kpc while \citet{mi20} get $> \sim 0.2 R_{\rm vir}$  
for the Milky Way-sized galaxies ($M_{\rm vir} \sim 10^{12} {\rm M}_\odot$ at $z=0$) for $0<z<2.4$.
\citet{an17} find the median maximum distance larger than $\sim 0.2 R_{\rm vir}$ for the halo mass
above $10^{11} {\rm M}_\odot$ but decreasing toward smaller masses with $\sim 0.1 R_{\rm vir}$ (close to 
the galaxy-scale) 
at $10^{10} {\rm M}_\odot$ for $0<z<3$. However, their $10^{10} {\rm M}_\odot$ halo (m10) has the present stellar mass 
of only $1.8\times 10^6 {\rm M}_\odot$, much smaller than the lowest mass $\sim 3 \times 10^{7} {\rm M}_\odot$ 
explored in this study. Nevertheless, gas recycling likely involves a wide range of spatial scales 
as suggested by these simulations. It should be also noted that the simulations by 
\citet{gr19} and \citet{mi20} include 
AGN feedback in addition to stellar feedback. AGN feedback, however, is considered to have 
significant impact primarily for massive halos and would not affect gas ejection in low-mass halos. 
To summarize, available simulation data suggest that the ejected gas typically travels well outside 
discs before re-accretion in low-mass halos, for which gas recycling plays an important role.
Finally, we just mention that the recycling gas may have already been detected as strong ${\rm C}_{\rm IV}$ absorbers 
against background quasars \citep{fo14}. Its suggested extent ranges from $\sim 0.1-0.5 R_{\rm vir}$ \, 
increasing with the halo mass \citep{ha22}, in overall agreement with the result from the 
above-mentioned simulations.

\begin{figure*}
	\includegraphics[width=\linewidth]{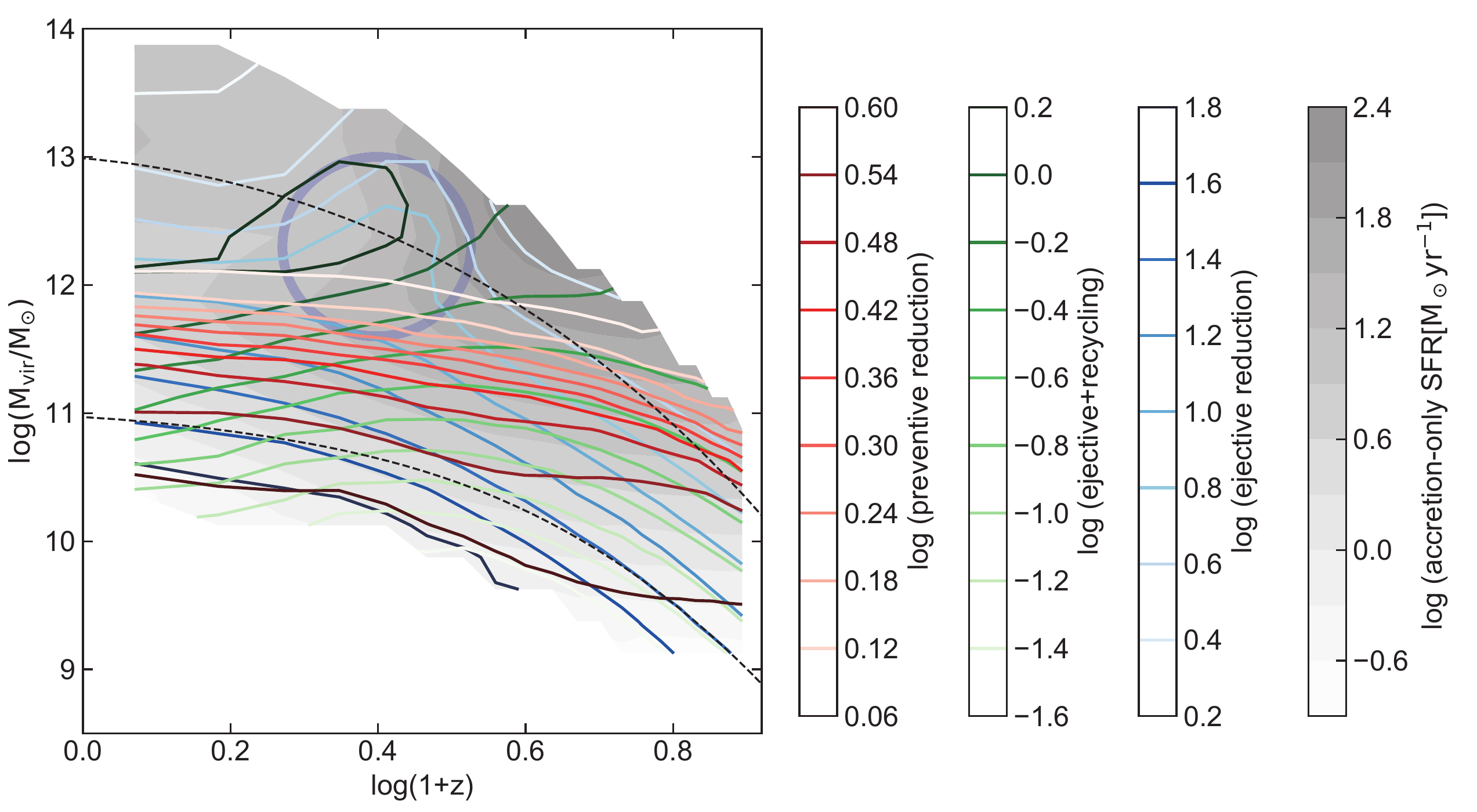}
    \caption{
	    The effect of feedback and recycling plotted on the $M_{\rm vir}-z$ plane.
	    Gray shades show the star formation rate for the accretion-only model.
	    Contours show the effect of each process on the accretion-only SFR.
	    Ejective (blue) and preventive (red) feedback reduce the SFR especially in low-mass halos.
	    Gas recycling partly remedies reduction of the SFR
	    caused by ejective feedback. Combined effect of these two processes 
	    is given in the form of the SFR relative to the accretion-only SFR (green). 
	    Two dashed lines running from upper left to lower right indicate the representative pass 
	    for high-mass and low-mass halos.
	    The blue circle indicates the domain where SFR is heavily suppressed by 
	    ejected feedback.
         }
\end{figure*}

Figure 16 visualizes the effect of each process on the star formation rate discussed above 
on the $M_{\rm vir}-z$ plane. The grey shading represents the SFR in the accretion-only model.
The suppression or enhancement of this SFR by each process is shown by coloured contours as indicated.
The evolution tracks of two models 
representative of high and low mass regimes
are plotted by dashed lines. 
Active star formation driven by the cold-mode accretion is readily recognized 
above ${\rm log M}_{\rm vir}[M_\odot]\sim11.5$  at $z\gtrsim2$ (log(1+$z$)$\gtrsim$0.5).
It is noted that a new domain 
appears for massive halos (blue circle) in contours for ejective feedback, where the SFR is more suppressed than in the surrounding region. 
 It is usually considered that supernova feedback has 
no significant effect in massive halos but this may not hold always especially in presence of
active star formation activity which massive galaxies experience under cold-mode accretion.

Detailed analysis reveals that this SFR suppression is caused by the strong gas ejection 
associated with the preceeding active star 
formation and leads to the small half-mass lookback time shown  in 
Figure 14 for $M_{\rm vir}=10^{13}{\rm M}_\odot$ as explained below. 
Figure 17 compares SFH and accretion history of the accretion-only model and the model to which 
only ejective feedback is added. Upper panels illustrates the relation of gas accretion rates and star formation.
Grey dashed line indicates the gas accretion rate in the accretion-only model.
This accretion history produces the SFH (grey solid line) that is a bit delayed from the accretion.
When ejective feedback is included (blue lines), net accretion rate (blue dashed line),
which is the difference between the accretion rate of the halo gas and the rate of gas ejection 
(blue dotted line), is reduced with respect to the 
accretion-only case.  It decreases almost to zero after the massive ejection associated with the initial 
starburst, leading to steeper drop in the SFR than in the accretion-only case.
Lower panel of Figure 17 shows how this feedback affects the growth of the stellar component.
The magenta dashed line indicates that ejective feedback is effective in reducing the accumulation of the ISM gas 
($M_{\rm g}$) especially in early times because of smaller halo masses and escape velocities.
What is notable here is a dip seen around 10 Gyr ago (log(1+$z$)$\sim$0.4), that is related to peak gas ejection shortly before.
The effect on the SFR (magenta solid line) is qualitatively similar to the change in $M_{\rm g}$.
This large dip delays the stellar mass growth by about 4 Gyr as shown by comparing grey and blue 
solid lines that indicates the fractional mass growth 
in agreement with the delay shown in Figure 14. 
It is interesting if this strong feedback process suggested for massive halos
is observed as gas outflow in high-redshift massive galaxies.

\begin{figure}
	\includegraphics[width=\linewidth]{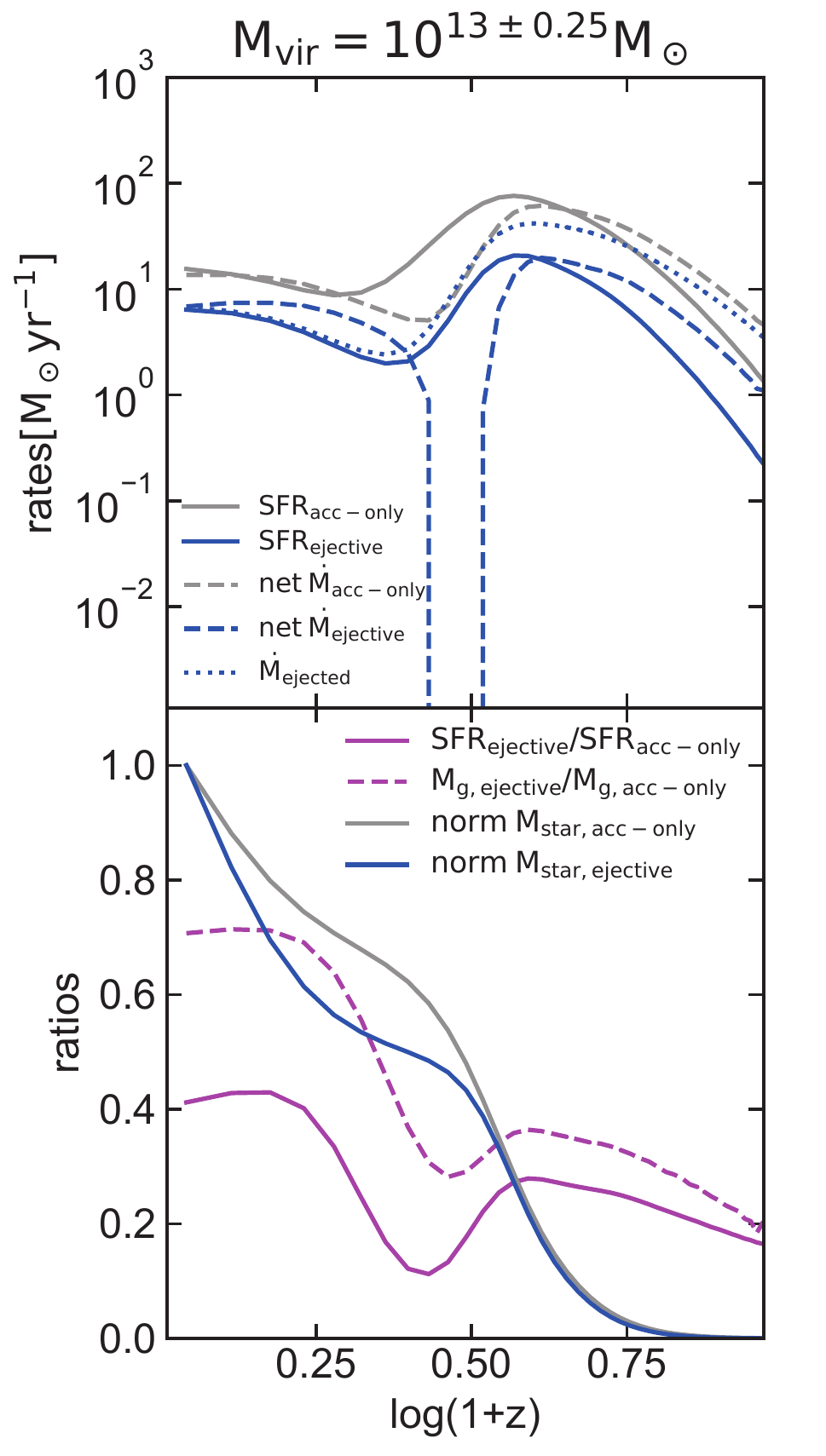}
    \caption{
	    The effects of ejective feedback on the evolution of the ISM and the stellar component
	    in halos of $M_{\rm vir}=10^{13} {\rm M}_\odot$ in the fiducial accretion scheme.
	    Upper panel: Rates of gas accretion, ejection and star formation. Grey and blue lines 
	    show the result for the accretion-only model and the model to which ejective feedback 
	    is added (corresponding to those two models plotted in Figure 12).
	    net $\dot{M}_{\rm ejective}$ is the accreion-only accretion rate (grey dashed line) 
	    minus the gas ejection rate (blue dotted line).
	    Lower panel: Relative reduction of the gas mass ($M_{\rm g}$) and SFR (magenta lines) 
	    caused by ejective feedback. Grey and blue lines indicate the fractional stellar 
	    mass growth in the two models.
         }
\end{figure}

\begin{figure}
	\includegraphics[width=\linewidth]{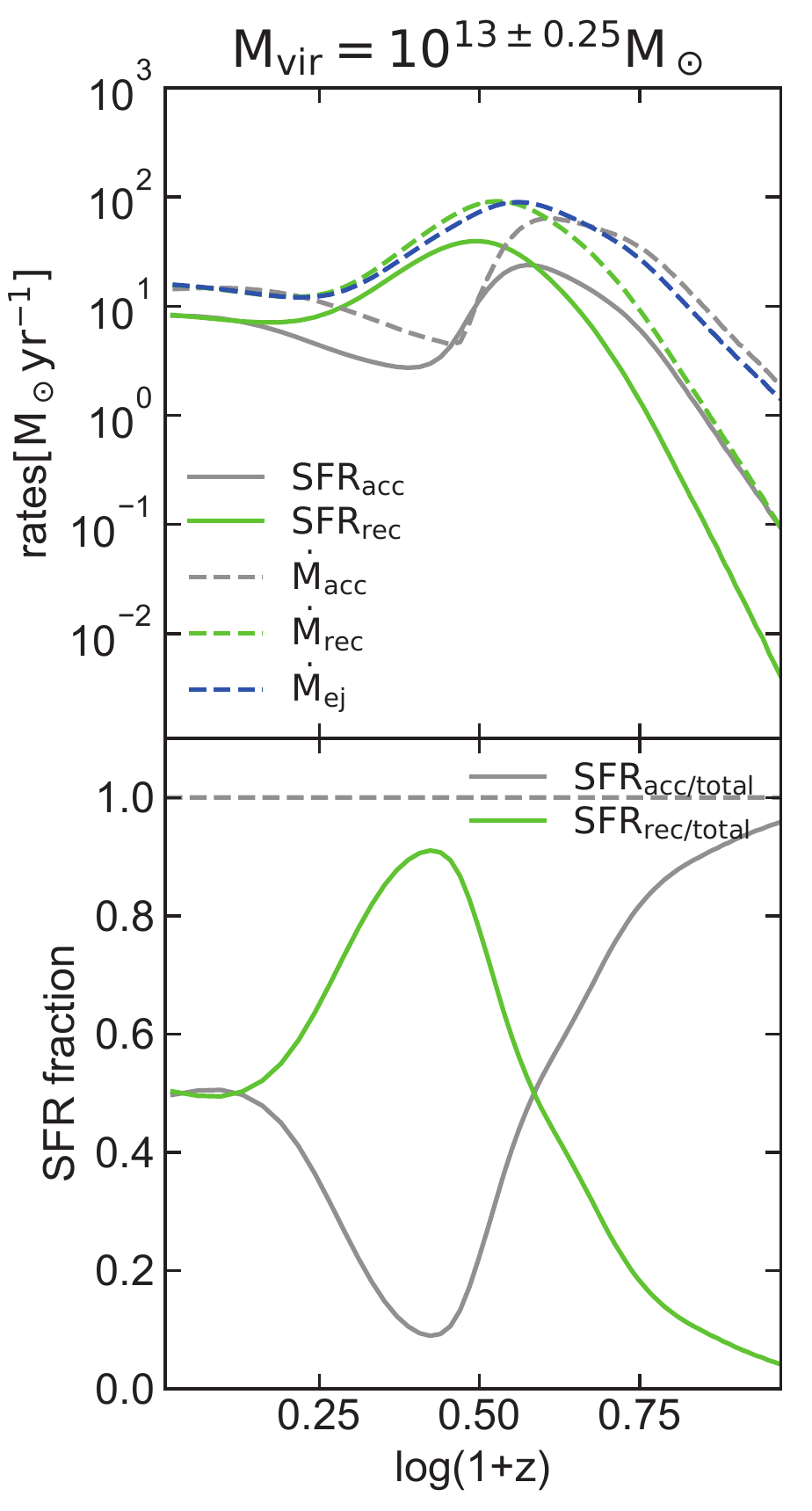}
    \caption{
	    Relative contribution of first-time gas accretion and recycling gas accretion in star formation
	    for the full model 
	    with $M_{\rm vir}=10^{13} {\rm M}_\odot$ in the fiducial scheme.
	    Upper panel: Breakdown of the SFR into two compoments with two different accretion routes.
	    Rates of first-time accretion, ejection and recycling accretion are plotted with dashed lines.
	    Lower panel: Fraction of each SFR component with respect to the total SFR.
         }
\end{figure}

Such a violent gas ejection also affects the gas recycing process in massive halos because the
reservoir of the ejected gas increases accordingly.
Figure 18 shows the relative contribition of the first-time gas accretion and the recycled accretion 
to the SFR in massive halos. This breakdown of instantaneous SFR was calculated as follows.
The SFR in the present model is calculated from the total ISM gas mass $M_{\rm gas}$ 
as described in Appendix A3. We added an attribute to this mass that specifies the mass fractions of two components 
originatng in the halo gas that accreted for the first time and in the recycling accretion 
from the ejected gas reservoir. At each timestep, we monitor the respective amounts of gas added by the halo gas accretion
(i.e., first-time accretion) and by the recycling, and update the attribute correspondingly.
The stars formed at each timestep inherit the attribute of the ISM gas at that timestep.
Implicit assumption underlying this method is that the accreted gas with different origins is 
mixed completely at each time. Testing this assumption necessitates numerical simulations with 
much finer temporal and spacial resolution than attainable at present.
We reserve this issue as a future task and take this assumption as a caveat at the moment.

Upper panel of Figure 18 shows that the initial increase in gas accretion from the halo (grey dashed line) 
triggers active star formation that is delayed a bit and peaks at ${\rm log(1+z)}\sim 0.6$ (grey solid line).
This starburst causes strong gas ejection with a little delay (blue dashed line).
Because of short recycling time implemented, gas recycling catches up with gas ejection immediately 
after the peak ejection (green dashed line). Recycling-driven star formation (green solid line) follows 
recycing gas accretion. This chain of events brings about significant time lag between two modes 
of star formation. Lower panel of Figure 18 indicates the fractions of two star formation modes.
Initial short period is dominated by the first-time gas accretion from the halo because 
there is not enough time for gas recycling to follow the gas ejection. Just after the peak 
${\rm SFR}_{\rm acc}$, gas recycling starts to become dominant route due to time lag.  
Although the contribution of recycling accretion diminishes with time, it remains a major source
for star formation in halos with this mass range. Because the recycling rate follows 
the ejection rate almost completely except intitial epoch where halo and galaxy masses are small,
 the recycling of the ejected ISM returns most of the ejected gas 
in massive halos and nearly recovers the SFR in the accretion-only case as shown in Figures 11 and 15. 
Recovery ability 
of this process diminishes toward lower halo masses.

\section{Discussion}

The present model adopting the fiducial scheme of gas accretion reproduces reasonably well 
several key features 
in the observed SFHs for a range of galaxy mass. There, however, remain issues to be clarified before 
we reach a more consistent picture for galactic star formation.
Although the reproduction of the quantitative detail is not the purpose of this study,
we here discuss several topics not touched upon in previous sections, 
hoping to get some hint for further study toward more comprehensive and quantitative 
understanding.

\subsection{Degeneracy of ejective and preventive feedback}

The importance of preventive feedback was suggested by the results discussed above.
We introduced this process by reducing the gas accretion rate to the disc.
Ejective feedback also helps suppress star formation by removing the interstellar gas.
These two processes operate effectively in 
similar domains in the $M_{\rm vir}-z$ plane (Figure 16), suggesting a possibility 
that these processes are degenerate.
Namely, a question arises whether ejective feedback with properly adjusted parameters 
can serve as a substitute for the required preventive feedback.

\begin{figure*}
	\includegraphics[width=\linewidth]{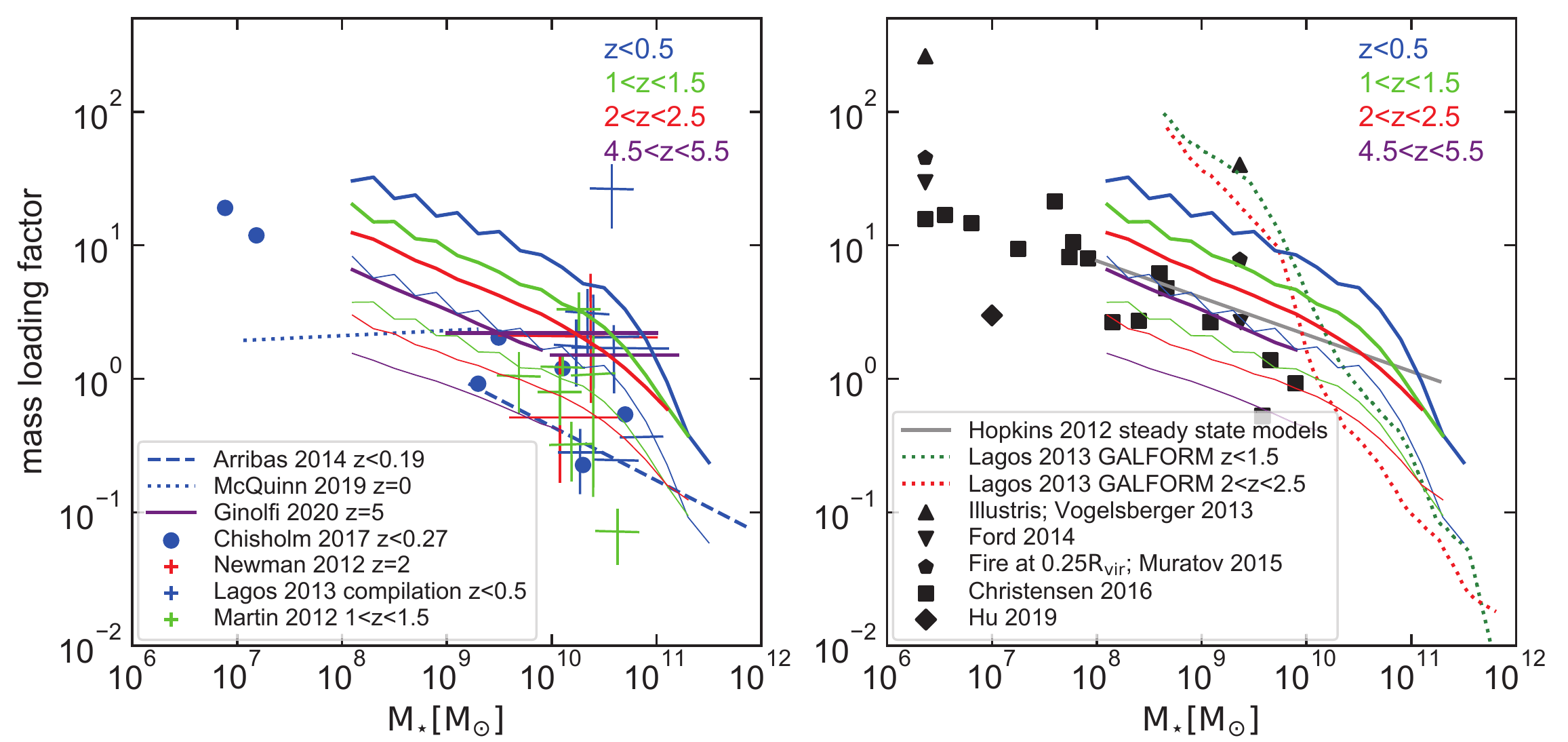}
    \caption{
	    Mass loading factors obtained in the present model compared with the observation (left) and 
	   previous theoretical studies (right).
	   Redshift is colour-coded as indicated.
	   Thick and thin solid lines indicate the models in the fiducial scheme with the feedback effciency
	   of 0.2 and 0.05, respectively.
	   Left panel: Observational results are shown by \citet{ar14}, \citet{ch17}, \citet{gi20}, 
	   \citet{ma12} and \citet{mc19}
	   (The observation by \citet{gi20} is two horizontal purple lines).
	   Blue crosses and short horizontal segments represent the results
	   by various authors 
	   for nearby galaxies (see Fig. 14 of \citet{la13} for details).
	   Right panel: 
	   The numerical simulation for steady state models by \citet{ho12} is indicated by the
	   grey solid line because no redshift is specified. 
	   The semi-analytic model (GALFORM) by \citet{la13} for $z<1.5$ and $2<z<2.5$ is indicated by 
	   green and red dotted lines, respectively.
	   Black symbols indicate the results by cosmological simulations for $z=0$ from  
	   \citet{vo13}, \citet{fo14}, \citet{mu15}, \citet{ch16} and \citet{hu19}.
         }
\end{figure*}

 To explore this possibility, we first turned off preventive feedback and 
increased the efficiency of ejective feedback 
up to unity (i.e., all the energy liberated by supernovae is used in ejecting the ISM).
For any choice in this range ($0.2 < \varepsilon_{\rm FB} <1$), the replacement by ejective feedback 
 exacerbates the poor fit to
the observed stellar-to-halo mass ratio. Matching the ratio for low mass halos inevitably leads to
 unacceptable underproduction of the stellar content for high mass halos.
This may be due to the steep $V_{\rm vir}^{-2}$ dependence of the energy-driven feedback adopted here
($V_{\rm vir}$ is the virial velocity of the halo).
We next tried the momentum feedback with the  $V_{\rm vir}^{-1}$ dependence.
This choice did not give a satisfactory agreement either.
This result means that raising the ejection efficiency above the fiducial value 
is not a promising replacement for preventive feedback.

This in turn casts a doubt that the efficiency of 20\% adopted for the fiducial 
ejective feedback may already be too large.
In order to check this point, we examined 
 the mass loading factor (i.e., the ratio of the mass ejection rate and the SFR, hereafter MLF) in the present model. Figure 19 compares our fiducial-scheme model with available 
 observations and other theoretical models.
The MLFs for the efficiency $\varepsilon_{\rm FB}=0.2$ roughly border the upper envelope for the observational data.
In the low mass domain where feedback has dominant effect, the semi-analytic model by \citet{la13}
and Illustris simulation by \citet{vo13}
give higher MLFs than our model while other cosmological simulations and the steady state models by \citet{ho12} show lower values.

Considering this result, we lowered the efficiency
 to $\varepsilon_{\rm FB}=0.05$.
This brings the model in better agreement with
the observation and the theoretical results. For this value, massive halos  
 suffer from negligible effect of ejective feedback.
 In this case, however, we need much stronger action of preventive feedback
that reduces the gas accretion rate to $\sim 5\%$ instead of the adopted 30\% in order to
match the stellar mass in low-mass galaxies. This leads to 
the contribution in the stellar mass reduction being comparable between two types of feedback while
the contribution of ejective feedback is about twice that by preventive feedback in the 
$\varepsilon_{\rm FB}=0.2$ case. 
It is not clear if such a strong preventive feedback is possible.
Because the observationally derived MLFs likely give only lower limits (but also see the 
discussion by \citet{mc19}),
the ejective efficiency of $20\%$ may not necessarily be unrealistic. 
Incidentally, \citet{du09} found that $\sim 25 \%$ of the supernova energy is required to 
reproduce the scaling relations of disc galaxies although this conclusion was derived for the shock-heating picture.
In any case, it seems robust that we need a significant contribution from preventive 
feedback to explain the stellar mass budget over the whole mass range.

\subsection{Comparison of the adopted preventive feedback to previous works}

The importance of prevented feedback has been stressed also from the chemical point of view
\citep[e.g.][]{mi15,ch16}.
It is well known that there is a tension between low efficiency of making stars and retaining 
enough metal in low-mass galaxies.
The model that relies solely on ejective feedback in reducing stellar 
masses produces a too steep mass-metallicity relation
 compared with the observation \citep[e.g.][]{to14,so15}. Preventive 
feedback can reduce the stellar mass in low mass galaxies with
their metallicity nearly untouched 
 \citep[e.g.][]{to14}.
For example, the cosmological simulation by \citet{ch16} that includes preventive feedback produces
the mass-metallicity relation at $z=0$ with the slope and normalization agreeing with the observation.


\begin{figure*}
	\includegraphics[width=\linewidth]{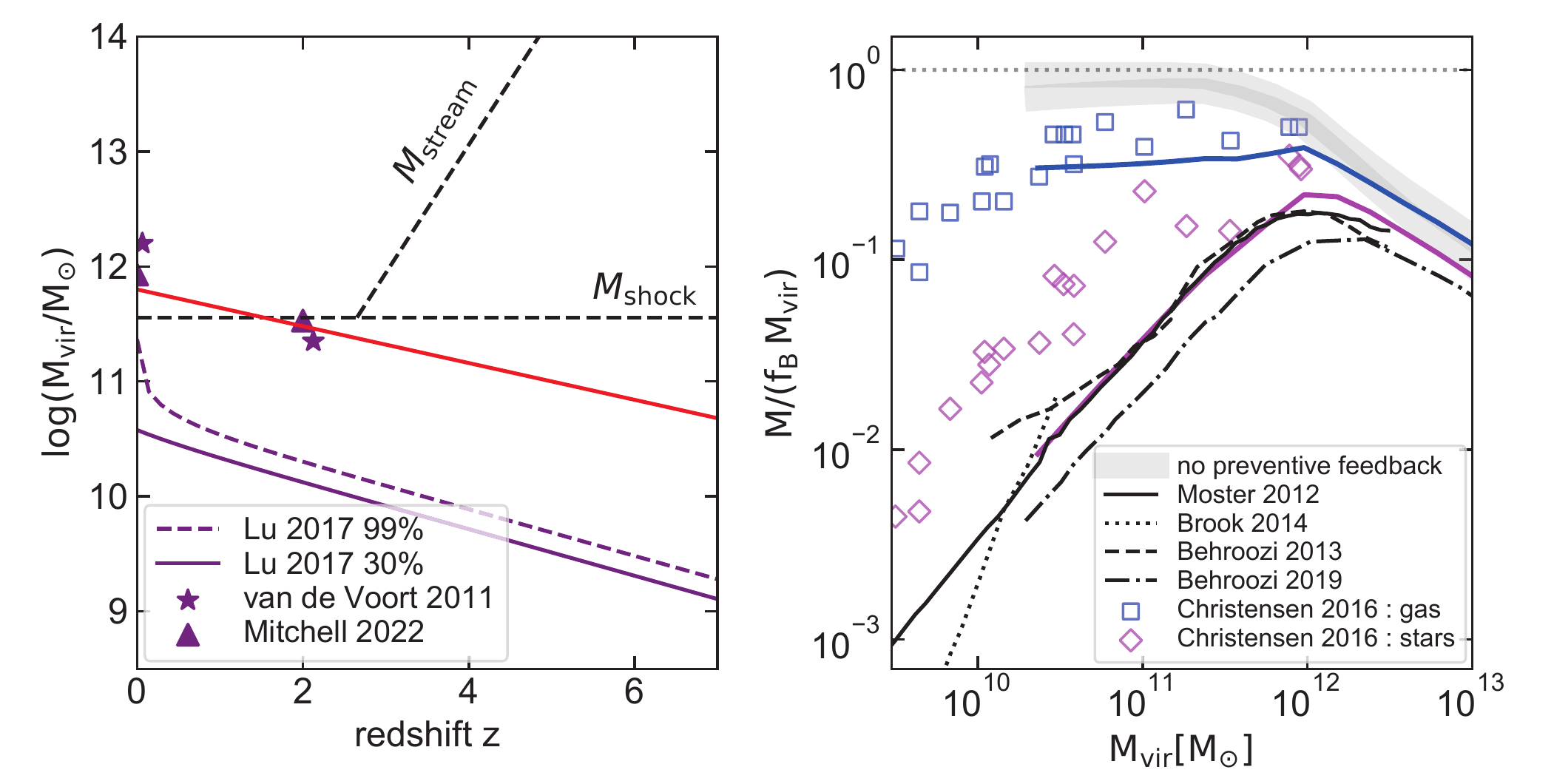}
    \caption{
	 Preventive feedback in this study compared with other works.
	  Left: The critical halo mass below which the gas accretion is prevented 
	  is indicated by the red line. Purple lines indicate the halo mass at which 
	  99$\%$ and 30$\%$ of the gas is accreted in the prescription used by \citet{lu17}.
	  Asterisks and triangles indicate the mass below which the gas accretion is suppressed
	  in the cosmological simulations by \citet{va11} and \citet{mi22}, respectively.
	  Right: Masses of stars and gas present in the disc at $z=0$ as a function of the halo 
	  virial mass. Masses are normalized by the cosmic baryon fraction times the virial mass.
	  Blue and magenta solid lines indicate the gaseous and stellar masses for the full model
	  in the fiducial scheme
	   while gray broad lines indicate the model in which 
	  no preventive feedback is included. Squares indicate the gas accreted to the $\it disc$ 
	  in the simulation by \citet{ch16} for direct comparison with the present model
	  in which the prevention is modelled as the reduction of accretion on to discs (not to halos).
	  Diamonds indicate the stellar mass in the disc from the same simulation.
	  Black lines denote the results obtained by the abundance matching technique by the 
	  indicated authors.
         }
\end{figure*}

We have seen above that preventive feedback is required also to reproduce the stellar-to-halo
mass ratio for low mass galaxies.
Nevertheless, preventive feedback is here introduced in an ad-hoc manner tuned for better reproduction
of the observational data. It is important to check the plausibility of the introduced 
prescription to describe this process. 
Because there is no strong constraint from the observation at the moment, we compare the outcome 
of our prescription to the cosmological simulations addressing this process.
The red line in the left panel of Figure 20 indicates the critical halo mass for preventive 
feedback adopted in this study.
Two asterisks indicate the halo mass below which the inclusion of {\it stellar} feedback 
reduces {\it the gas accretion rate on to halos} in the simulation by \citet{va11}.
Simulations by \citet{co18} and \citet{wr20} give similar results.
Two triangles denote the mass below which the prevented gas starts to dominate the total baryon mass 
budget in EAGLE simulation performed by \citet{mi22}.
Our prescription agrees with these cosmological simulations at least up to $z \sim 2$.
The FIRE-2 cosmological simulation by \citet{ho18} and the SAM by \citet{pa20} also suggest the critical mass
for preventive feedback caused by supernovae that decreases with redshift.
Their critical masses lie in the range $10^{11}-10^{12} {\rm M}_\odot$ in agreement with our prescription and 
other cosmological simulations.
Analytical prescription considered in \citet{da12} adopts  systematically lower efficiency than our choice 
over the redshift range $0<z<2$ and applies to lower halo masses.
Nevertheless, their recipe dictates stronger feedback for higher redshifts and less massive halos,
which is qualitatively similar to our recipe.

Right panel of Figure 20 shows that the decrease in the gas accretion 
due to preventive feedback is around $30 \sim 40 \% $ for 
$M_{\rm vir} < 10^{12} {\rm M}_\odot$ in our fiducial-scheme model. This diminishment is comparable to the one 
obtained in the cosmological simulation by \citet{ch16}.
For the stellar mass, our model gives much smaller values but in agreement with the empirical estimates for the 
stellar-to-halo mass ratios.
This fact may lend some support
 to the recipe adopted in the present work.

\citet{lu17} parametrized the strength of preventive 
feedback as a function of halo mass and redshift by fitting their semi-analytic model 
to the stellar mass function and 
the mass-metallicity relation of 
the satellite galaxies of Milky Way.
Dashed and solid purple lines in the left panel indicate the halo mass 
for which the accretion to the halo is reduced to respectively $99\%$ and $30\%$
with respect to $f_{\rm bar} \dot M_{\rm vir}$. These masses are 
much lower than our recipe and other studies mentioned above,
 rather in agreement with 
the boundary for UV background suppression given by \citet{gn00} and \citet{ok08} (see Figure 1).
Their model thus seems to imply that the background radiation field is sufficient 
in preventing gas accretion in the Milky Way system.
By contrast, other studies suggest a close link between ejective and preventive feedback.
For example, \citet{op10}, based on the numerical simulations, argue that the suppression of gas accretion by 
 supernova wind energy and momentum has a strong effect for low mass galaxies. 
\citet{va11} and \citet{mu15} also appear to support this picture.

The suggested dominance of preventive feedback for low mass halos 
may demand significant revision of the cold accretion picture for the low-mass regime.
Possible interactions between the outgoing gas ejecta and the inflowing cold gas may change 
the physical state of the halo gas drastically.
In other words, preventive feedback by definition invalidates the assumption of the universal baryon 
fraction in parent halos and/or the gas accretion rate (on to discs) calculated from either
the free-fall time or the radiative cooling time.
Nevertheless, we could gauge
 the required suppression of gas accretion on to galaxies relative to 
 the accretion rate calculated from the universal baryon fraction and appropriate inflow timescales 
 (also see Appendix, section A6).

\subsection{Missing processes}

The present study demonstrates the ability of simple isolated models in reproducing 
the observed diversity in galactic star formation history. This success is attained by
assuming the cold accretion regime of halo gas aquisition and cooperation of
ejective feedback, preventive feedback and gas recycling. This study reveals at the same time 
several shortcomings of these idealized models. The most notable disagreement with the 
observational data is that the models underpredict SFR at high redshifts. 
At low redshifts, massive galaxies show higher SFRs than observed.
Here we discuss several processes which are neglected in our models but 
may resolve or alleviate these problems. Specifically, we touch on here AGN feedback, galaxy mergers
and wind transfer.

The most promising process to reduce high SFR in massive galaxies in recent epochs is
feedback from AGNs.
It has been considered to be one of important processes that potentially affect 
the galaxy evolution and has been taken into account in many theoretical studies
\citep[e.g.][]{si07,vo13,we18}.
These studies (both the SAMs and cosmological simulations) succeeded in explaining the quenching 
of massive galaxies by introducing AGN feedback
\citep[e.g.][]{ka00,cr06,de10,lu14}.
Quenching mechanisms can act in various forms.
For example, radio mode feedback can act as preventive feedback \citep[e.g,][]{ke09,zi20}
and may lead to 'halo quenching' as discussed by \citet{lu14}.
The accumulation of AGNs on way from the blue cloud to the red sequence for late type 
galaxies \citep[e.g.][]{sc10,po17} may suggest AGN-quenching for star forming galaxies from the observational 
viewpoint.  
\citet{ma19} report that the inclusion of AGNs (LINERs and Seyfert 2 galaxies) 
 into observational analysis leads to the turn-over of the 
MS at high masses.
On the other hand, \citet{de10}, using a semi-analytic model, argue that the necessity of AGN feedback is changed by 
various assumptions adopted for gas cooling. 
Recent cosmological simulations \citep[e.g.][]{co18,ne19,wr20} give more quantitative results for the effect of AGN feedback 
on gas accretion rates. For example, \citet{wr20} indicate that 
  stellar feedback has  much stronger preventative effect than AGN feedback 
(factor $\sim4$ versus $20\sim30\%$).
They infer that this difference is due partially to the former acting preferentially in 
 low mass halos that have shallow gravitational potentials 
while the latter operates in massive halos having deep potential wells.
Our result also suggests that preventive feedback  is more important in low mass halos
in line with these simulation result (e.g., Figure 16).
Probably, one of the most relevant results to the current issue is provided by
the EAGLE simulation \citep{ka17} which shows  that AGN feedback at low redshifts improves 
the agreement of the simulation with the observed evolution of the cosmic SFR density (CSFRD)
(see their Figure 10).
Although CSFRD is the integrated quantity over all galaxy populations, this result hints at 
important role of AGN feedback in quenching massive galaxies in recent epochs. 
This result suggests that inclusion of AGN feedback could help cure unrealistic upturn in SFR
in low-redshift massive galaxies seen in the present fiducial scheme.

	Regrettably, most observational works conducted to date \citep[e.g.][]{sc10,mc19} give only circumstantial 
evidence for AGN quenching. More direct information could be provided by measuring the mass loading factor
for AGN-driven outflows. For example, \citet{qi21} calculated the ratio of outflow rate and  black hole 
accretion rate for massive elliptical galaxies. If some scaling relation between this ratio and the BH 
accretion rate or BH mass is established by extending such analysis to larger samples including 
late-type galaxies, 
it could be combined with the
BH growth model to provide better theoretical insight into the importance of AGN feedback for different galaxy masses and 
cosmological epochs. 

On the other hand, overly inactive star formation in model galaxies at high redshift could be 
alleviated by galaxy mergers which are expected to be frequent at early cosmological epochs. 
Along with AGN feedback, galaxy mergers have also been considered as an important process 
affecting mainly massive galaxies. They can provide a promising route for morphological
transformation and star formation quenching, both processes possibly leading to the 
formation of early-type (elliptical) galaxies. 
One concern with this scenario is that mergers may not quench star formation 
completely. For example, the simulation by \citet{sp05} indicates that the aid from AGN feedback is necessary 
to expell the residual ISM completely outside merging galaxies.
Despite this caveat, it seems evident that galaxy mergers (especially major ones) have non-negligible 
effects on the evolution of galaxy populations.
Including mergers is difficult in our modelling by construction.
Therefore, the present model should be, in the strict sense, regarded as mainly describing evolution of disc-dominated galaxies
that have probably suffered little from the effects of major galaxy mergers in their lives.
Although the merger-driven starbursts in disc galaxies are thus not handled in our current model, 
it seems reasonable to expect that frequent mergers in early times would enhance 
star formation activity by perturbing young gas-rich galactic discs.

Recent cosmological simulations \citep[e.g.][]{an17,gr19,mi20} point out the importance of intergalactic transfer 
of material such as 'wind transfer' in addition to mergers of galaxies themselves.
These simulations suggest comparable contribution of this process to other processes such as 
gas recycling (fountain flow) and fresh gas accretion.
Our model is unable to predict how  these extragalactic ingredients affect galaxy evolution.
What is naturally expected is that the idealized concept of hot and cold modes of gas accretion 
 depicted in Figure 1 may be significantly modified (in addition to possible modification 
 posed by preventive feedback mentioned in section 6.2).  
 Currently, the definition of these newly introduced processes itself varies among different authors 
and mutual comparison of simulation results is not straightforward.
 It is challenging but worthwhile to forge various processes occuring on halo scales into
 a unified picture as the next target.

\section{conclusions}

We constructed a simple model for the evolution of galactic stellar contents 
in order to investigate how the star formation history is affected by gas accretion process
from host halos.
We consider three different schemes of gas accretion.
The shock-heating scheme assumes that the gas gathered by halos experiences heating by 
shock waves to the virial temperature and subsequent radiative cooling induces gas accretion.
According to the stability analysis of shock waves, low mass halos experience 
rapid free-fall accretion of unheated gas due to instability irrespective of redshift. The flat 
scheme takes this picture.
Finally, the fiducial scheme assumes efficient accretion dirven by filamentary  
streaming of cold gas through the shock-heated hot gas in massive 
halos at high redshift, in addition to the cold-mode accretion induced by shock instability
in the low mass domain.
In addition to gas accretion, the model 
 includes two types of feedback; ejective feedback by supernova energy and 
 preventive feedback
that hinders some portion of the halo gas from accreting on to galactic discs.
Re-accretion (recycling) of the interstellar medium ejected by supernovae 
is also taken into account.

The fiducial scheme reproduces
several key features of the  
observational results regarding the star formation activity in massive galaxies.
The early occurrence of the star formation peak and the strong diminishment of the 
star formation rate after that are reproduced at least qualitatively.
Moreover, the slope of the high-mass domain in the SFR-$M_\star$ relation (MS) decreases with time
in agreement with observations.
	On the other hand, the flat and shock-heating schemes predict similar SFHs in massive halos 
	that  qualitatively contradict observations. Namely, SFRs increase with 
	time nearly monotonically and the high-mass end of MS shows steeper slopes toward
	recent epochs.
The transition from the cold to hot modes of halo gas accretion, which roughly 
marks peak star formation activity, is an important piece 
in shaping the star formation history in massive halos.
Retardation of this transition in the fiducial scheme caused by inflow of cold gas streams in high-mass high-redshift 
 domain leads to a larger gap 
in accretion timescale and therefore a more drastic drop in accretion rate and SFR.

In contrast to massive halos, low-mass halos evolve under rapid gas accretion 
	in all accretion schemes. In the presence of only accretion, they experience 
	active star formation in early epochs reflecting growing history of their parent halos, 
	leading to too massive stellar components and old ages at present time.
Comparable contribution of supernova feedback and  preventive 
feedback seems to be required to reduce their stellar masses to the observed level.
Recycling of the ejected interstellar medium with the timescale of $\sim 1$ Gyr 
plays an essential role in producing monotonically increasing SFR and 
thus making these galaxies young enough to match the observation.
In the fiducial scheme, this change in SFHs of low mass halos and the early strong quenching
in massive halos lead to the downsizing galaxy formation as observed.

\section*{Acknowledgements}

We thank Camilla Pacifici, Gerg\"{o} Popping, Kate Whitaker for providing their data.
Data compilations for other studies used in this paper were made by using 
 the online \textsc{WebPlotDigitizer} code.\footnote{\url{https://automeris.io/WebPlotDigitizer/}}.

\vspace{16pt}

\noindent{\bf DATA AVAILABILITY}

\vspace{6pt}
\noindent{The data underlying this article will be shared on reasonable request to the corresponding author.}





\appendix

\section{Numerical Methods}

\subsection{Growth of dark matter halos}

We follow the parametrization of the mass accretion history of a single halo 
by \citet{be13} (APPENDIX H), which gives an excellent fit to Bolshoi \citep{kl11} and Multidark \citep{ri13}
cosmological simulations. This parametrization also solves the problem in Wechsler's 
parametrization \citep{we02} adopted in previous works 
\citep{no18,no20} that mass accretion histories can cross at high redshifts.
Although this problem is unlikely to have serious effects for the redshift range we concern,
we choose to adopt a more consistent treatment taken in \citet{be13}.

For the concentration parameter of the halo needed in the calculation of the gas 
cooling time, we follow the result of cosmological simulations
for WMAP5 cosmology summarized by \citet{ma08}.
Namely, the  halo concentration at present is given by ,

\begin{equation}
	  \log C_{\rm vir} = 0.971 - 0.094 \log(M_{\rm vir}/[10^{12}h^{-1}{\rm M}_{\odot}])
\end{equation}

Following \citet{bu01}, we assume that the concentration parameter
evolves as

\begin{equation}
	 c(z) = \min \left[ K \frac{1+z_{\rm c}}{1+z} , K \right]
\end{equation}
with $K=3.7$. Here $z_{\rm c}$ is the collapse redshift of the halo calculated
once the present concentration $c(0)=C_{\rm vir}$ is specified as
\begin{equation}
	z_{\rm c} = c(0)/K
\end{equation}
A minimum value on $c(z)$ is imposed following the result by \citet{zh03}
to mimic the early growing phase of the halo with nearly constant concentration.

\subsection{Gas accretion rates}

The calculation of gas accretion rate requires the knowledge of 
dynamical free-fall time and cooling time.

The free-fall time, $t_{\rm ff}$, is

\begin{equation}
	t_{\rm ff} =  \sqrt {{3\pi} \over {32G\rho_{\rm vir}}}
\end{equation}
where
\begin{equation}
	\rho_{\rm vir} = {{3M_{\rm vir}} \over {4\pi R_{\rm vir}^3}}
\end{equation}
is the average halo density within the virial radius.

In calculating the cooling time, the halo gas is assumed to be in collisional ionization equilibrium with $T=T_{\rm vir}$.
The cooling time, $t_{\rm cool}$, is then given by
\begin{equation}
	t_{\rm cool} = \frac{3}{2} \mu m_{\rm p} \frac{kT_{\rm vir}}
	{\rho_{\rm hot}\Lambda(Z_{\rm gas},T_{\rm vir})}
	\frac{\mu_{\rm e}^2}{\mu_{\rm e}-1}
\end{equation}
Here, $\rho_{\rm hot}$ is the density of virialized gas
and given by the total mass density at the virial mass multiplied by
the cosmic baryon fraction $f_{\rm bar}$ which is assumed to be 0.17.
 $\Lambda$ is the normalized cooling function for an ionized gas in collisional ionization equilibrium
 given by \citet{su93} that depends on the gas metallicity $Z_{\rm gas}$ and temperature $T_{\rm vir}$ for helium mass abundance of 0.25.
No photoionization background is considered here.

We choose a constant gas metallicity, $Z_{\rm gas} = 0.01$ based on the following consideration.
\citet{pr03} report that their sample of $\sim 100$ damped Ly$\alpha$
(DLA) systems show the increasing metallicity with the time, 
from $Z \sim 10^{-2}$ at $z=4$ to $\sim 10^{-1}$ at $z=1$.
If these DLAs probe innermost regions of halos, the metallicity at the outer halos 
should be smaller than the reported values.
\citet{sy03} report that the metallicity of the intergalactic medium
depends on the overdensity but shows no evidence for evolution.
The metallicity increases with the overdensity $\delta$, reaching $\sim 10^{-2}$ at 
$\delta=10^2$. Because the virial radius is determined so that the mean overdensity
inside the virial radius is $\sim 200$, the metallicity at the virial radius 
at which we calculate the cooling time is likely to be smaller than $10^{-2}$.
Examination of the effect of varying halo gas metallicity revealed that 
the results do not change appreciably for $Z<10^{-2}$. We thus take $Z=10^{-2}$ as the 
fiducial value. We do not take into account the metal enrichment from star formation
which is unlikely to affect the global gas accretion rate but the innermost region.

The mass accretion rate of the halo gas is calculated as follows. In each time step, 
the increase of the halo mass during that time step is calculated based on the halo growth
model. The gas mass acquired during this period is the halo mass increment
multiplied by the cosmic baryon fraction. This gas is assumed to be distributed 
spherically in accordance with the NFW profile having the concentration parameter as 
given above. We can calculate the free-fall time and 
the radiative cooling time for the virial temperature as stated above. Depending on the 
current location of the halo in the $M_{\rm vir} - z$ plane (Fig.1) and the adopted scheme, we can determine 
what timescale, i.e., either the free fall time or the radiative cooling time,  we should apply. 

This procedure, performed in this form, leads to discontinuous accretion rates when 
the halo crosses the borders for different accretion domains. 
The division into different accretion domains shown in Fig.1 is idealized treatment
while the transition through a border reported in cosmological simulations is
actually more gradual. In order to alleviate 
this difficulty, we apply smoothing as follows. We consider a rectangular smoothing window which has widths
$wz$ in redshift and  $wm$ in logarithmic virial mass. Then we evaluate the relative area 
of each accretion domain included in this window. The newly added gas mass is allotted
to each accretion domain according to its fraction and assigned with respective accretion 
timescale.  In the models reported here, window widths of $wz = wm =0.5$ were taken and 
each direction was divided into 100 meshes to calculate relative contribution of each domain.
This choice was found to retain essential features of the model behaviour while erasing the discontinuity 
effectively.

\subsection{Star formation}

\begin{figure*}
	\includegraphics[width=1.0\linewidth]{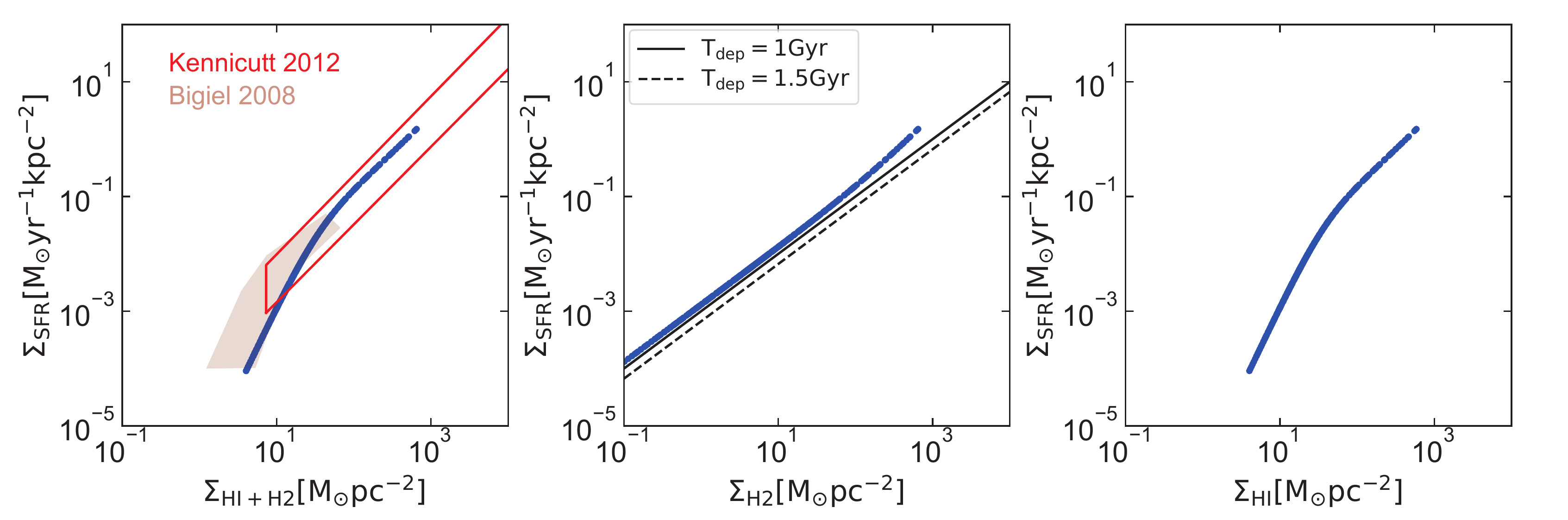}
    \caption{
	    Kennicutt-Schmidt law in the fuducial scheme (blue) campared with the observational data.
	    Left: The star formation rate density as a function of the total gas 
	    surface density. The dashed line is a eye-fit to the relation by \citet{ke12},
	    whereas the shaded region approximately represents the distribution
	    of galaxies given in \citet{bi08}.
	    Centre: The same as left panel but as a function of the molecular gas
	    surface density. 
	    Nearly linear relation for the models is noted.
	    Solid and dashed lines denote the relation for 
	    the gas depletion time of 1 Gyr and 1.5 Gyr, respectively.
	    The model depletion time is consistent with the observed value 
	    derived by \citet{ta13}.
	    Right: The same as centre panel, but as a function of the neutral gas
	    surface density. Turn-down at low gas density is noted.
         }
\end{figure*}


We follow \citet{bl06} in implementing star formation.
The gas accreted from the halo is assumed to form a gaseous disc and replenish the interstellar medium 
(ISM), from which stars are formed. 

Star formation rate density, $\Sigma_{\rm SFR}$, is given by

\begin{equation}
	 { { \Sigma_{\rm SFR} } \over {(\rm M_{\odot} pc^{-2} Gyr^{-1})} } = 
	 { {\epsilon_{\rm SF}} \over {(\rm  Gyr^{-1})}} {{\Sigma_{\rm HCN}}
	  \over {(\rm M_{\odot} pc^{-2})}}  
\end{equation}
where $\epsilon_{\rm SF} = 13 {\rm Gyr^{-1}}$
  \citep{ga04,wu05}.
Here, the surface density of the dense molecular gas is given by,

\begin{equation}
	   \Sigma_{\rm HCN} =  f_{\rm HCN} \Sigma_{\rm mol} 
\end{equation}
  where $\Sigma_{\rm mol}$ is the surface density of total molecular gas and
  the ratio $f_{\rm HCN}$ depends on the molecular gas density \citep{bl06} as

\begin{equation}
	     f_{\rm HCN} = 0.1 \left[ 1 + { {\Sigma_{\rm mol}} \over
	     {(200 \rm M_{\odot} pc^{-2})}} \right] 
\end{equation}

  Total molecular density is derived from the surface density of the cold gas, $\Sigma_{\rm cold}$, as

\begin{equation}
	  \Sigma_{\rm mol} = f_{\rm mol} \Sigma_{\rm cold} 
\end{equation}
  where

\begin{equation}
	    f_{\rm mol} = {  { R_{\rm mol}} \over { 1 + R_{\rm mol}} }    
\end{equation}
  and

\begin{equation}
	   R_{\rm mol} = \left({{P/k} \over {4.3 \times 10^4} }  \right)^{0.92} 
\end{equation}
  which we take from \citet{bl06} and is valid for 
	   $10^{-1}<\Sigma_{\rm mol}/\Sigma_{\rm HI}<10^2$, where the atomic gas surface density
	   is $\Sigma_{\rm HI}=\Sigma_{\rm cold}-\Sigma_{\rm mol}$.

  Here, $P$ is the gas pressure at the disc plane and $k$ stands for the Boltzmann constant.

  Following the dynamical argument by \citet{el93}, the gas pressure in the disc is given by

\begin{equation}
	    P = {\pi \over 2} G \Sigma_{\rm cold} 
	    \left[ \Sigma_{\rm cold} + ( {{\sigma_{\rm cold}}
	     \over {\sigma_{\rm star}}}) \Sigma_{\rm star} \right]   
\end{equation}
   where $\Sigma_{\rm star}$ is the surface density of stars.
  Surface densities for the cold gas and stars are calculated as  
	    $\Sigma_{\rm cold} \equiv M_{\rm gas}/(\pi R_{\rm disc}^2)$ and 
	    $\Sigma_{\rm star} \equiv M_{\rm star}/(\pi R_{\rm disc}^2)$, respectively,
	    where the disc radius is set to be $\lambda R_{\rm vir}$.
	    The spin parameter is taken to be $\lambda=0.06$.
Ratio of the gaseous and stellar velocity dispersions is assumed to 
be $\sigma_{\rm cold}/\sigma_{\rm star}=0.1$ following \citet{du09}.

Instant recycling approximation is used to calculate the mass returned to the ISM
due to evolution of massive stars with the returned fraction of $R=0.44$ 
, which is an approximate value for the results reported by \citet{vi16}
for the Chabrier IMF \citep{ch03}
 for a wide range in metallicity.

 The adopted star formation recipe succeeds in reproducing the observed relation between 
 gas surface densities and the star formation rate surface density as shown in Figure A1.

\subsection{Supernova feedback}

Massive stars provide feedback to the surrounding interstellar medium 
when they explode as supernovae, 
with the result that some portion of the ISM is ejected from the galaxy.
The ejected mass  is calculated as

\begin{equation}
	 \Delta M_{\rm ej} = \left[{{2 \epsilon_{\rm FB} E_{\rm SN} \eta_{\rm SN}}
	  \over {V_e^2}} \right] \Delta M_{\rm star}
\end{equation}
where $\Delta M_{\rm star}$ is the mass of stars formed in the current time step,
$\epsilon_{\rm FB}$ is the feedback efficiency,
$E_{\rm SN}$ is the energy released by one supernova, and $\eta_{\rm SM}$ is the
number of supernovae produced per  one solar mass of stars formed at that step.
We take $E_{\rm SN}=10^{51}$ erg and $\eta_{\rm SN}=8.3 \times 10^{-3}$ following \citet{du09}.
For feedback efficiency we adopt $\epsilon_{\rm FB}=0.2$ as the fiducial one.

The escape velocity is calculated as

\begin{equation}
	V_{\rm e}  = \sqrt {2 G M_{\rm vir}/R_{\rm vir}}  
\end{equation}

We call this model of feedback  'ejective feedback' to distinguish it from  'preventive feedback' 
described below.
We also tried the momentum driven prescription instead of the above energy-driven one as mentioned
in the text. In this case, equation (A14) is replaced by

\begin{equation}
	 \Delta M_{\rm ej} = \left[{{2 \epsilon_{\rm FB} p_{\rm SN} \eta_{\rm SN}}
	  \over {V_e}} \right] \Delta M_{\rm star}
\end{equation}
 
Here, $p_{\rm SN}=3\times10^4{\rm M}_\odot {\rm km s}^{-1}$ is the momentum
produced by one SN taken from \citet{mu05}.

\subsection{Reincorporation (Recycling) of ejected gas}

The ultimate fate of the ejected gas is highly uncertain. We assume that some portion 
of the ejected gas falls back and returns to the galactic disc for further star formation.
We use the parametrization by \citet{mi15} for the timescale of reincorporation (recycling) 
of the ejected gas as

\begin{equation}
	t_{\rm rec}  = 0.52 \times 10^{9} {\rm yr} \times (1+z)^{-0.32}
	\left( M_{\rm vir} \over 10^{12}{\rm M}_\odot  \right)   ^{-0.45}
\end{equation}
where the virial mass is given in units of the solar mass.

Although this is obtained as the best fit for their equilibrium model, it was 
found to reproduce the observation best  for the present evolutionary model 
among many trials with different parameter values.

The mass recycled at each time step is calculated as 

\begin{equation}
	M_{\rm rec}  = \left( M_{\rm ej} / t_{\rm rec}\right)  \Delta t
\end{equation}
where $M_{\rm ej}$ is the total mass of ejected material present at the current time step
(i.e., the reservoir of the ejected material) 
and $\Delta t$ is the (constant) length of a single time step.
It should be noted that $M_{\rm ej}$ increases by stellar ejective feedback 
and decreases by recycling. Namely, it is updated at each time step as

\begin{equation}
	M_{\rm ej} = M_{\rm ej} + \Delta M_{\rm ej} - M_{\rm rec}
\end{equation}

\subsection{Preventive feedback}

To mimic preventive feedback in a simple form, 
we reduce the gas accretion rate to the disc calculated as above to  the level

$$ f_{\rm p} = f_{\rm pr}+(1-f_{\rm pr})(y+\frac{\pi}{2}) $$
\noindent{where }

$$ y={\rm arctan}[3({\rm log} M_{\rm vir}-{\rm log} M_{\rm pr})/w_{\rm pr}] $$

\noindent{Here} $f_{\rm pr}$ is the parameter controlling the strength of feedback, $M_{\rm pr}$
is the critical mass below which  feedback is significant, and $w_{\rm pr}$ controls
how suddenly  feedback gets strong with the decreasing halo mass. 
$M_{\rm pr}$ is a function of redshift and given by

$${\rm log} M_{\rm pr}=M_0+z(M_5-M_0)/5 $$
\noindent{where} $M_0$ and $M_5$ are the critical mass at $z=0$ and 5, respectively.
We take $f_{\rm pr}$=0.3, $M_0$=11.8, $M_5$=11.0, and $w_{\rm pr}$=0.5 as the default values unless stated otherwise.
For this choice, $f_{\rm p} \sim 0.65$ at the critical mass.
The original accretion rate is multiplied by the calculated factor $f_{\rm p}$ to model preventive feedback.


\bsp	
\label{lastpage}
\end{document}